# Endothelial Cell-specific Loss of Breast Cancer Susceptibility Gene 2 Exacerbates Atherosclerosis

David C. R. Michels[1], Sepideh Nikfarjam[1,2], Berk Rasheed[1,2], Margi Patel[1], Shuhan Bu[1], Mehroz Ehsan[1], Hien C. Nguyen[1,2], Aman Singh[1], Biao Feng[3], John McGuire[1], Robert Gros[4], Jefferson C. Frisbee[1], Krishna K. Singh[†,1,2]

Department of [1]Medical Biophysics, [2]Anatomy and Cell Biology, [3]Pathology and Laboratory Medicine,[4]Physiology and Pharmacology, Schulich School of Medicine and Dentistry, University of Western Ontario, London, ON, Canada.

Author emails: dmichel6@uwo.ca; snikfarj@uwo.ca; brashee@uwo.ca; mpate292@uwo.ca; sbu4@uwo.ca; mehsan2023@meds.uwo.ca; hnguy29@uwo.ca; asing945@uwo.ca; biao.feng@schulich.uwo.ca; john.mcguire@schulich.uwo.ca; rgros@uwo.ca; jfrisbee@uwo.ca; krishna.singh@uwo.ca

**Running title:** Role of Breast Cancer Gene 2 in Atherosclerosis.

[†]**To whom correspondence should be addressed:** Krishna K. Singh, Department of Medical Biophysics, Schulich School of Medicine and Dentistry, University of Western Ontario, 1151 Richmond St. N., London, ON, N6A 5C1; Phone: (519) 661-2111 x 80542 (Office) x 85683 (Lab); Email: krishna.singh@uwo.ca.

**Disclosure of Conflicts of Interest:** None.

**Abstract**

**Background:** The BReast CAncer type 2 susceptibility protein (BRCA2) responds to DNA damage by participating in homology-directed repair. BRCA2 deficiency culminates in defective DNA damage repair (DDR) that when prolonged leads to the accumulation of DNA damage causing cancer or apoptosis. DNA damage represents a common basis for diseases characterized by increased oxidative stress, including but not limited to cancer and cardiovascular diseases (CVDs). Oxidative stress promotes DNA damage and apoptosis and is a common mechanism through which cardiovascular risk factors lead to endothelial dysfunction and atherosclerosis. Herein, we present new evidence that endothelial BRCA2 plays a protective role against atherosclerosis under hypercholesterolemic stress.

**Methods:** We first successfully generated and characterized endothelial cell (EC)-specific BRCA2 knockout ($BRCA2^{endo}$) mice. Next, to study the effect of EC-specific BRCA2-loss in atherosclerosis, we generated and characterized $BRCA2^{endo}$ mice on apolipoprotein E null background ($ApoE^{-/-}$), fed them with high-fat diet (HFD) and evaluated atherosclerosis. Finally, RNA-seq analysis was performed on RNAs extracted from aortas of HFD-fed EC-specific BRCA2-deficient $ApoE^{-/-}$ and BRCA2-intact $ApoE^{-/-}$ mice.

**Results:** Baseline phenotyping of $BRCA2^{endo}$ mice did not show any adverse effects in terms of DNA damage and apoptosis as well as cardiac and metabolic function. However, using HFD-fed apolipoprotein E knockout ($ApoE^{-/-}$) background, we demonstrated that EC-specific loss of BRCA2 resulted in aortic plaque deposition and splenomegaly. Comparison of RNA sequencing data from aortas of EC-specific BRCA2-deficient $ApoE^{-/-}$ and BRCA2-intact $ApoE^{-/-}$ mice revealed a total of 530 significantly differentially expressed genes with Protein Folding Response and Lipid Metabolism as the most affected pathways.

**Conclusion:** This study provides foundational knowledge regarding BRCA2 status and function in the cardiovascular system, and highlights the potential of BRCA2 as a novel therapeutic target in prevention and treatment of atherosclerosis. Our data indicate that BRCA2 mutation carriers may be at a previously unrecognized risk of atherosclerosis in addition to breast and ovarian cancer.

**List of Abbreviations**

Ach: acetylcholine

ADAM17: ADAM metallopeptidase domain 17

ApoE: apolipoprotein E

BRCA2: BReast CAncer type 2 susceptibility gene

BSA: bovine serum albumin

CVD: cardiovascular disease

DAB: 3,3'-Diaminobenzidine

DAVID: Database for Annotation, Visualization and Integrated Discovery

DBD: DNA-binding domain

DDR: DNA damage repair

DEG: differentially expressed gene

DSB: double-strand break

EC: endothelial cell

GAPDH: glyceraldehyde 3-phosphate dehydrogenase

GO: gene ontology

H&E: haematoxylin and eosin

HDL: high-density lipoprotein

HFD: high-fat diet

HO-1: heme oxygenase-1

HRP: horseradish peroxidase

IPA: Ingenuity Pathway Analysis

LDL: low-density lipoprotein

lncRNA: long non-coding RNA

LVEDD: left ventricle end-diastolic diameter

LVEF: left ventricle ejection fraction

LVESD: left ventricle end-systolic diameter

LVFS: left ventricle fractional shortening

M/F: male and female

NLS: nuclear localization sequence

NLRP3: NLR family pyrin domain containing 3

NRF2: nuclear factor erythroid 2-related factor 2

OB: oligonucleotide binding domain

oxLDL: oxidized low-density lipoprotein

PALB2: partner and localizer of BRCA2

PE: phenylephrine

RIPA: radioimmunoprecipitation assay buffer

SDS-PAGE: sodium dodecyl-sulfate polyacrylamide gel electrophoresis

SNORD15B: small nucleolar RNA, C/D Box 15B

SNP: sodium nitroprusside

SNPs: single nucleotide polymorphisms

snRNA: small nuclear RNA

SREBF2: sterol regulatory element-binding transcription factor 2

SREBP2: sterol regulatory element-binding protein 2

SYCP1: synaptonemal complex protein 1

SW/BW: spleen weight to body weight ratio

TG: triglyceride

TGFBR2: transforming growth factor-β receptor 2

TUNEL: terminal deoxynucleotidyl transferase (TdT) dUTP nick-end labeling assay

VE-cad: vascular endothelial cadherin

WT: wildtype

**Novelty and Significance**

**What Is Known?**

- The tumor suppressor BRCA2 plays a critical role in the repair of damaged DNA and maintenance of genomic integrity.
- Mutations in *BRCA2* gene predisposes carriers to breast and ovarian cancers.
- Increased incidence of cardiovascular disease is reported in breast cancer patients.
- Endothelial dysfunction plays an important role in development of cardiovascular disease, such as atherosclerosis.
- Oxidized-LDL treatment exacerbates endothelial dysfunction in BRCA2-deficient endothelial cells in vitro.

**What New Information Does This Article Contribute?**

- In the absence of stress, endothelial cell-specific loss of BRCA2 results in no grossly apparent cardiovascular phenotype.
- Endothelial cell-specific loss of BRCA2 exacerbates atherosclerotic plaque deposition in high-fat diet-fed $ApoE^{null}$ mice.
- Endothelial cell-specific loss of BRCA2 causes splenomegaly in high-fat diet-fed $ApoE^{null}$ mice.
- Protein folding response and lipid metabolism are the most affected pathways in the aorta of endothelial cell-specific BRCA2-deficient high-fat diet-fed $ApoE^{null}$ mice.

- These data collectively highlight a novel role of BRCA2 as a gatekeeper of endothelial cell function in atherosclerosis, and suggest evaluating endothelial function and atherosclerosis in *BRCA2* mutation carriers.
- *BRCA2* mutation carriers may be at a previously unrecognized risk of atherosclerosis in addition to breast and ovarian cancer.

## Introduction

Atherosclerosis is a chronic immunoinflammatory condition in which endothelial dysfunction plays an early and permissive role.[1-3] Atherosclerosis-associated cardiovascular diseases (CVDs) represent a leading cause of death globally,[9] and they progress via cardio-oncological factors including DNA damage,[9-15] inflammation,[16] apoptosis,[10, 12, 17, 18] and cellular dysfunction.[19] Subendothelial retention of low-density lipoprotein (LDL) and its oxidative modification represent the initial events during atherogenesis.[4] Oxidized LDL (oxLDL) activates endothelial cells (ECs), which in turn mediate the rolling and adhesion of blood leukocytes to the intima layer via chemotaxis.[4] OxLDL-induced macrophage activation leads to the release of proinflammatory cytokines and proteolytic enzymes involved in matrix degradation, atherosclerotic plaque destabilization, and ultimately plaque rupture.[4] OxLDL-induced oxidative stress promotes DNA damage, EC dysfunction and apoptosis.[5-8] DNA damage, which occurs in all cell-types in the atherosclerotic milieu including ECs, vascular smooth muscle cells and macrophages[12, 14], exacerbates the oxidative stress-induced endothelial dysfunction[5, 12, 21] and is positively correlated with the disease severity.[9] Numerous studies have demonstrated the accumulation of DNA damage in atherosclerotic plaques either as genomic alterations or DNA adducts.[21-23]

The BReast CAncer susceptibility gene 2 (*BRCA2*) is classified both as tumor suppressor and caretaker based on its known function in homology-directed repair of DNA double-stranded breaks (DSBs) and regulation of cell cycle and transcription.[24-26] Germline mutations in *BRCA2* are highly penetrant[27, 28] and predispose women to an increased lifetime risk of breast and ovarian cancers by 69% and 17%, respectively.[29] The *BRCA2* gene encompasses 27 exons and encodes a 384 kDa protein.[30] Exon 11 of *BRCA2* comprises approximately 60% of the gene's coding sequences[31] and contains highly conserved RAD51-binding BRC repeats.[30] During homologous recombination repair, the BRC repeats facilitate recruitment of RAD51 to DSB sites[32, 33]. *BRCA2* mutations that hinder RAD51 recruitment[37], reduce DNA damage repair (DDR) capacity, and result in DNA damage accumulation[38] and cellular dysregulation[39], eventually leading to inflammation, apoptosis or cancer.[38].[5, 25, 39, 40] Given that accumulating evidence demonstrate the presence of DNA damage, inflammation and apoptosis in the pathogenesis of CVDs, [8, 41-44] it is likely that BRCA2 plays a significant role in the development of CVDs such as atherosclerosis.[42, 44-47] Human *BRCA2* resides on the chromosome 13q12.3, a region associated with CVDs.[55-58] Furthermore, a genetic analysis identified single nucleotide polymorphisms (SNPs) in *BRCA2* that were associated with increased risk of coronary artery disease[59] and enhanced plasma lipid levels.[59, 60] Additionally, lipid and metabolite deregulation were reported in *BRCA2*-mutant humans.[61, 62] In endothelium, DNA damage accompanied by lipid deregulation and apoptosis are the central pathways for atherosclerosis.[63, 64] Several studies indicated a causative role for BRCA2 in CVDs.[21, 40, 44, 47-54] OxLDL-induced endothelial dysfunction and apoptosis were exacerbated in BRCA2-deficient ECs in vitro.[5] Murine models with cardiomyocyte-specific loss of *BRCA2* were more susceptible to cardiotoxicity induced by genotoxic agents.[40] In humans, BRCA2 mutations were associated with a 2-fold increased risk of diabetes[50] and 37% increased risk of all-cause mortality among breast cancer patients, the latter of which was mainly due to CVDs.[52] Accordingly, excess non-neoplastic deaths were reported in *BRCA2*-mutant individuals.[51] Given the significant role of BRCA2 in genome homeostasis and regulation of cellular function and to further investigate the role of BRCA2 in the cardiovascular system, we aimed to assess the effects of loss of EC-specific BRCA2 in atherosclerosis. Herein, we report for the first time that loss of endothelial BRCA2 accelerates the progression of atherosclerosis. From a translational standpoint, our observations highlight a previously unrecognized role of BRCA2 as the gatekeeper of endothelial function and suggest that individuals carrying *BRCA2* mutations may be at a previously unrecognized susceptibility to develop atherosclerosis particularly in the presence of risk factors.

## Results

### Generation of Endothelial Cell-specific *BRCA2* Knockout Mice

To evaluate the basal expression level of BRCA2, we first measured BRCA2 transcript levels in different organs of WT mice and observed the highest expression levels in testis and ovary followed by spleen, aorta, and liver with minimal expression in heart (Fig. 1A). Given that all of these organs are highly vascularized, to directly elucidate the role of BRCA2 in vascular function and to circumvent the embryonic lethality associated with systemic loss of *BRCA2*[68], we applied the Cre-loxP system to delete *BRCA2* specifically in the endothelium.[21, 69, 70] Conditional inactivation of endothelial BRCA2 was achieved by crossing mice homozygous for the exon 11 floxed *BRCA2* allele ($BRCA2^{fl/fl}$) with hemizygous mice expressing Cre recombinase under the control of VE-cadherin (VE-$cad^{tg/-}$) promoter (Fig. 1B, C).[69, 70] Mice of genotypes VE-$Cre^{-/-}$;$BRCA2^{fl/fl}$, VE-$Cre^{-/-}$;$BRCA2^{fl/wt}$, VE-$Cre^{-/-}$;$BRCA2^{wt/wt}$, and VE-$Cre^{tg/-}$;$BRCA2^{WT/WT}$ are denoted as $BRCA2^{WT}$. Genotype VE-$Cre^{Tg/-}$;$BRCA2^{fl/WT}$ is denoted as $BRCA2^{het}$ and genotype VE-$Cre^{Tg/-}$;$BRCA2^{fl/fl}$ is denoted as $BRCA2^{endo}$ mice (Fig. 1B). PCR products using genomic DNA and primers B012/013 and B014/015 flanking the LoxP sites verified upstream *BRCA2* exon 11 and downstream LoxP sites, respectively (Fig. 1C, D).

ECs comprise ~30% of the lung cells[71]. Accordingly, we first measured BRCA2 expression level in the lungs of $BRCA2^{WT}$, $BRCA2^{het}$ and $BRCA2^{endo}$ mice. Using a *BRCA2* exon 11-specific primer, we observed a dose-dependent reduction of BRCA2 transcript in the lung tissues (Fig. 2A), which was further confirmed at protein level (Fig. 2B). We further confirmed EC-specific deletion of *BRCA2* by immunohistochemistry of the aortic and lung sections, where we observed reduced BRCA2 expression in the endothelium of $BRCA2^{endo}$ mice in comparison with the $BRCA2^{WT}$ (Fig. 2C). To confirm whether exon 11-deficient BRCA2 transcript ($BRCA2^{\Delta 11}$) is expressed and stable, we performed qPCR on aortic ECs isolated from $BRCA2^{WT}$ and $BRCA2^{endo}$ mice using primers spanning exon 10 to exon 14. Our data showed the presence of $BRCA2^{\Delta 11}$ transcript only in the aortic endothelium of $BRCA2^{endo}$ mice (Fig. 2D). Using N-terminal-specific BRCA2 antibody, we were able to detect a full-length WT BRCA2 protein in the aorta of $BRCA2^{WT}$ mice, but no truncated forms of BRCA2 protein was detected in the aorta of $BRCA2^{endo}$ mice. This indicates that although $BRCA2^{\Delta 11}$ transcript was present, it was either not translated or $BRCA2^{\Delta 11}$ protein was unstable after translation (Fig. 2E). Our finding is in line with a previous report where $BRCA2^{\Delta 11}$ protein was not detected in a tissue-specific knockout mice which were generated using the same $BRCA2^{fl/fl}$ line.[40]

**Endothelial Cell-*s*pecific Deletion of *BRCA2* is Dispensable at Baseline**

Understanding the function of endothelial BRCA2 and the impact of its deficiency within the context of cardiovascular pathophysiology requires comprehensive examination of gross and cardiovascular metrics at baseline. In our examinations, male and female (M/F) $BRCA2^{endo}$ and $BRCA2^{het}$ mice displayed no obvious phenotype and were born in expected Mendelian ratio, indicating the absence of embryonic lethality observed in systemic BRCA2 knockout (Fig. 3A).[68] Weight at 8-10 weeks was similar across the genotypes (Fig. 3B). Exon 11 of *BRCA2* encodes for RAD51-binding BRC repeats[30] that physically interacts with and facilitates RAD51 recruitment to DSB sites.[32, 33] Inability of BRCA2 to bind to RAD51 leads to perturbed DDR, thereby permitting DNA damage accumulation, which may eventually lead to apoptosis.[38] Accordingly, we next evaluated the effect of loss of *BRCA2* on DNA damage accumulation and repair by measuring

γH2AX and RAD51 foci, respectively, in the aorta, lung and liver of $BRCA2^{WT}$, $BRCA2^{het}$ and $BRCA2^{endo}$ mice. Our immunohistochemical analysis showed no difference in γH2AX-foci, indicating the absence of DNA damage and DDR activation at baseline (Fig. 3C). Immunohistochemical data was further confirmed by γH2AX immunoblotting performed on total protein extracted from lung tissues of $BRCA2^{WT}$ and $BRCA2^{endo}$ mice (Fig. 3D). Given that loss of *BRCA2* is associated with p53-mediated apoptosis,[5, 40, 72] we measured p53 protein expression in the lungs of $BRCA2^{WT}$ and $BRCA2^{endo}$ mice and did not observe any difference between the groups (Fig. 3D). The absence of apoptosis was further confirmed by TUNEL staining, in the aorta, lung and liver of $BRCA2^{WT}$, $BRCA2^{het}$ and $BRCA2^{endo}$ mice (Fig. 3E, F and Supp. Fig. 1).

Next, echocardiography was performed to assess baseline effect of loss of endothelial *BRCA2* on cardiac function. Analysis of results demonstrated comparable ejection fraction and fractional shortening, indicating similar cardiac function across M/F $BRCA2^{WT}$, $BRCA2^{het}$ and $BRCA2^{endo}$ mice at baseline (Fig. 4A-C). Next, metabolic parameters including $O_2$ and $CO_2$ volume, respiratory exchange, energy expenditure, food consumption, water intake, total activity, sleep time and body weight were measured at baseline in 8-10 weeks old male $BRCA2^{WT}$ and $BRCA2^{endo}$ mice. Our analysis showed similar profile for all the metabolic parameters between the $BRCA2^{WT}$ and $BRCA2^{endo}$ mice (Supp. Fig. 2A-I). Overall, all animals in our study were viable and fertile, and displayed no molecular, cardiac or metabolic abnormality at baseline. These data indicate that loss of *BRCA2* is dispensable at baseline.

**Endothelial Cell-specific Loss of *BRCA2* Accelerates Atherosclerosis in *ApoE*$^{-/-}$ Mice**

Conditional inactivation of *BRCA2* in endothelium of $ApoE^{-/-}$ mice was achieved as demonstrated in Fig. 5A and B. These mice were also born in the expected Mendelian ratio and displayed no difference in weight or other obvious phenotypic features at 10 weeks of age (Fig. 5C, D). We next sought to investigate the effect of loss of endothelial *BRCA2* on HFD-induced atherosclerosis in double knockout ($BRCA2^{het}$;$ApoE^{-/-}$ and $BRCA2^{endo}$;$ApoE^{-/-}$) mice. Body weight did not significantly change following 2, 4, 6, 8, 10 and 12 weeks of HFD between the groups (Fig. 5E). We evaluated the extent of atherosclerosis by measuring plaque buildup area in the aortic arch. Plaque burden was measured as the percentage of regional total vessel area and visualized after Oil Red-O staining (Supp. Fig. 3A). At 8 weeks post-HFD, $BRCA2^{het}$;$ApoE^{-/-}$ and $BRCA2^{endo}$;$ApoE^{-/-}$ mice exhibited respectively 92% and 100% significantly greater plaque burden compared to WT controls (Fig. 6A, B). Similarly, at 12 weeks post-HFD, $BRCA2^{endo}$;$ApoE^{-/-}$ mice significantly increased plaque burden (by 81%) was observed in the aortic arch (Fig. 6C). These findings were further confirmed by staining of whole aorta, where we observed higher plaque deposition in the aortic arch of $BRCA2^{endo}$;$ApoE^{-/-}$ animals in comparison with the WT mice (Fig. 6D). Moreover, similar trend of genotype-dependent increase of plaque burden was observed after 16 weeks of HFD (Fig. 6E). Given that loss of *BRCA2* is known to promote a pro-inflammatory milieu by increasing macrophage infiltration,[39] we next measured the extent of inflammation by F4/80 staining of the aortic root. Our data revealed increased inflammation in $BRCA2^{endo}$;$ApoE^{-/-}$ in comparison with $BRCA2^{WT}$;$ApoE^{-/-}$ mice (Fig. 6F). Taken together, these findings support our hypothesis that loss of endothelial *BRCA2* exacerbates atherosclerosis and implicate the protective role of BRCA2 throughout the progression of CVD.

**Endothelial Cell-specific Loss of *BRCA2* Does Not Affect Plasma Lipid Profile in Atherosclerotic Mice**

BRCA2 has been associated with lipid and metabolite deregulation in humans[59-62]. Therefore, we conducted plasma lipid profiling of male $BRCA2^{endo}$;$ApoE^{-/-}$ and $BRCA2^{WT}$;$ApoE^{-/-}$ mice both maintained on HFD for 16 weeks. To our surprise, we did not observe any appreciable changes in total LDL, HDL, and cholesterol levels (Table 1).

**Endothelial Cell-specific Loss of *BRCA2* Causes Splenomegaly in Atherosclerotic Mice**

A relatively large spleen was observed in HFD BRCA2$^{endo}$;ApoE$^{-/-}$ mice as shown by significantly higher spleen weight to body weight (SW/BW) ratio (Fig. 7A, B). H&E staining of the spleen sections demonstrated enlarged splenic nodules due to increased cellularity in the white pulp region, indicating splenic lymphoid hyperplasia (**Fig. 7C**). Importantly, splenic sections of HFD BRCA2$^{WT}$;ApoE$^{-/-}$ and WT control mice showed a clearly defined splenic nodules of white pulp (**Fig. 7C**). Measurements of SW/BW of BRCA2$^{WT}$ and BRCA2$^{endo}$ mice at baseline did not reveal any differences, which rules out the effect of endothelial *BRCA2* deficiency in splenic lymphoid hyperplasia. (Supp. Fig. 3B). BRCA2 is also known to regulate the expression of heme oxygenase-1 (HO-1) via binding to the nuclear factor erythroid 2-related factor 2 (NRF2)[73, 74], and contribute to prevention of vascular inflammation. Therefore, we measured the expression level of HO-1 in the spleen of HFD BRCA2$^{het}$;ApoE$^{-/-}$ and BRCA2$^{endo}$;ApoE$^{-/-}$ mice and observed a significantly reduced level of HO-1 (Fig. 7D). However, loss of endothelial *BRCA2* alone was not sufficient to induce HO-1 expression in the absence of HFD (Supp. Fig. 3C).

**Endothelial Cell-specific Loss of *BRCA2* Differentially Regulates Protein Folding Response in Atherosclerotic Mice**

To identify the underlying molecular mechanisms involved in increased plaque burden and splenomegaly in HFD BRCA2$^{endo}$;ApoE$^{-/-}$ mice, we performed RNA sequencing and gene expression network analyses on aortic total RNA and identified a total of 530 significantly DEGs with 188 up- and 342 down-regulated genes (cut-off: ±1.5 and $p < 0.05$) (Supp. Table 1, Fig. 8A).

The top 20 up- and down-regulated genes were listed in Table 2 and 3, respectively. *SYCP1* (**s**ynaptonemal complex protein 1; 415-fold) and *SNORD15B* (small Nucleolar RNA, C/D Box 15B also known as U15; 193-fold) were the most up- and down-regulated DEGs. DEGs were annotated in the IPA database, to identify the enriched canonical pathways (Fig. 8B). The most important enriched pathways included Unfolded Protein Response, Regulation of Epithelial-Mesenchymal Transition Pathway and NRF2-mediated Oxidative Stress Response (Fig. 8B). Analysis of activation state of molecules by IPA revealed that several genes involved in tumorigenesis such as *TP53*, *TP63* and *TP73* were significantly inactivated while factors known to promote endothelial dysfunction such as *TGFBR2* (transforming growth factor-β receptor 2),[75] ADAM17 (ADAM metallopeptidase domain 17)[76] and SREBF2 (sterol regulatory element-binding transcription factor 2)[77] were activated (Fig. 8C). We further categorized the 530 DEGs based on gene ontology (GO) by using the Database for Annotation, Visualization and Integrated Discovery (DAVID) into Molecular Function, Biological Processes and Cellular Components (Fig. 8D). These results highlight dysregulation of genes involved in Protein Folding and Cholesterol Biosynthetic Process, which are associated with loss of endothelial *BRCA2* in $ApoE^{-/-}$ mice.

## Discussion

Cancer and CVDs are the major causes of mortality worldwide.[8, 44] Importantly, there are several fundamental similarities in the pathophysiological mechanisms contributing to cancer and CVDs, which have led to the emerging field of "Cardiovascular-Oncology"- a driving force for nascent research and development of precision medicine.[8, 44] Numerous associations have been suggested between CVDs, particularly atherosclerosis, and breast cancer[51, 59, 60, 78-81]; however, the underlying physiological drivers of these manifestations are yet to be uncovered. While germline mutations in *BRCA1* and *BRCA2* increase the susceptibility to breast cancer[82], accumulating evidence indicate a protective role for these genes in CVDs.[21, 40, 44, 47-53] Questions concerning the biological functions of the proteins encoded by *BRCA1* and *BRCA2* have dominated the field since the genes were identified through the analysis of families at high risk for breast and ovarian cancer. BRCA1 and BRCA2 are very large proteins, widely expressed in different tissues during S and G2 phases, and localize to nucleus. *BRCA1*[47] and *BRCA2* are both expressed differentially in the various organs of mice (Fig. 1A). Mice harboring cardiomyocyte- or EC-specific *BRCA1* mutations were more susceptible to developing cardiovascular dysfunction under imposed stresses.[21, 40, 47-49, 53] Loss of *BRCA1* exacerbated endothelial dysfunction and atherosclerosis progression by specifically modulating the DDR pathway in ECs.[21] Accordingly, a gain of *BRCA1* function had a protective effect on ECs.[21] Taken together, while these reports indicate a causative role for BRCA in the development of CVD in humans,[5, 21, 40, 44, 47-53, 59, 80, 83-86] there is only one study that failed to find an association between BRCA and CVD.[87] In all studies, however, there remains an important gap in knowledge and that is the precise prevalence of mutations in *BRCA* and/or haploinsufficiency among cardiovascular patients which makes it difficult to generalize the current knowledge about the association between BRCA and CVDs.

We recently reported that oxLDL exacerbated endothelial dysfunction and apoptosis in BRCA2-deficient ECs.[5] It is important to note that oxLDL treatment, which inhibited BRCA2 expression, did not affect the expression of BRCA1.[5] This is mainly due to the fact that BRCA2 protein is structurally completely distinct from BRCA1. Throughout homologous recombination, BRCA2 interacts with different protein partners, and thus plays a non-redundant role in different stages of genome protection. Additionally, BRCA2 has a different rate of cancer predisposition, lifetime risk and estimated frequency, and differs in the types and characteristics of cancers compared to BRCA1.[82, 88, 89] Unlike BRCA1, BRCA2 is associated with lipid regulation and inflammation in humans.[39, 54, 59, 90] *BRCA2* resides on the human chromosome 13q12.3, a region linked to CVDs.[55-58] At a functional level, BRCA1 is omnipresent with a wealth of biochemical data describing multiprotein interactions.[91] BRCA2, on the other hand, is mainly known to exert a more direct role to promote DDR.[92, 93] Therefore, the differential roles played by BRCA1 and BRCA2 warrants independent studies investigating their functions in cardiovascular biology. Several evidence demonstrated heightened rates of early mortality, irrespective of cancer,[51]among *BRCA2* mutation carriers. This provides the impetus for deeper understanding of common pathophysiological mechanisms. Beyond the established rate of cancer in *BRCA2* mutation carriers,[94, 95] these patients exhibited increased rate of CVDs[78, 79, 96].

The health of endothelium is a determinant of cardiovascular well-being and is altered in atherosclerosis to a condition referred to as 'endothelial dysfunction' characterized by inflammation and apoptosis. The occurrence of these complications in the vascular endothelium promotes atherosclerosis[97, 98] and further sensitize the endothelium to atherogenic factors including DNA damage, oxidative stress, and inflammation.[99, 100] Atherogenesis synergistically promotes DNA damage,[5, 9, 12-14] further essentializing BRCA2-mediated homologous recombination repair[101] and cell cycle maintenance functions.[102-104] Loss of *BRCA2* function mutations are therefore of primary interest for comprehensive evaluation in models of atherosclerosis. DNA DSB repair is primarily facilitated by the exon 11 of BRCA2 that contains conserved BRC repeats.[105] Systemic *BRCA2* ablation results in embryonic lethality in mice prior to day 9.5 of gestation due to its critical role in cell proliferation during early embryogenesis.[106] Endothelium-specific deletion of BRCA2 using the Cre-LoxP technology and VE-cadherin promoter allows to overcome this obstacle.[40] Since mice embryos have already reached late gastrulation[107] by the time VE-cadherin activity becomes detectable at day 7.5 of embryogenesis,[108] it leads to successful generation of viable $BRCA2^{endo}$ mice. Experimental mice were restricted to hemizygous Cre to limit the possibility of Cre toxicity[44] and untargeted DNA alterations.[109, 110]

Aberrant signalling by ECs to cardiomyocytes may result in impaired cardiac contractility.[111, 112] Accordingly, echocardiography was performed to assess cardiac function and rule out early signs of coronary artery disease,[47] cardiomyopathy,[101, 113] mitral valve dysfunction,[114] and hypertension.[115] Analysis of LVEF and LVFS revealed no difference in baseline cardiac function among $BRCA2^{WT}$, $BRCA2^{het}$ and $BRCA2^{endo}$ mice (Fig. 4 B and C). Similarly, no genotype-dependent differences were observed in metabolic function. Taken together, these data demonstrate that endothelial-specific deletion of *BRCA2* does not impact DNA damage and repair or cardiovascular function and may thus be considered dispensable at baseline. These findings are congruent with previous findings where cardiomyocyte-[47] or EC-specific loss of *BRCA1*[21] or cardiomyocyte-specific loss of *BRCA2*[40] did not display any remarkable molecular or cardiac abnormality at baseline. Lack of baseline phenotype in these mice provided us with the opportunity to explore the role of endothelial BRCA2 using an atherosclerotic mouse model.

HFD-fed $ApoE^{-/-}$ mice are well-established models of atherosclerosis characterized by hypercholesterolemia.[116-118] Accordingly, EC-specific BRCA2-deficient model was generated on $ApoE^{-/-}$ background. Vascular regions prone to atherosclerotic lesions are branches and segments exposed to high turbulence and low shear stress[17, 119] such as the aortic arch. These regions are more vulnerable to endothelial injury, and thereby are the prime candidate to evaluate disease progression. In our study, we found significantly increased plaque burden in $BRCA2^{het}$;$ApoE^{-/-}$ and $BRCA2^{endo}$;$ApoE^{-/-}$ mice following 8, 12, and 16 weeks of HFD. Moreover, a strong trend demonstrating increased aortic arch plaque burden in $BRCA2^{het}$;$ApoE^{-/-}$ vs $BRCA2^{WT}$;$ApoE^{-/-}$ and in $BRCA2^{endo}$;$ApoE^{-/-}$ vs $BRCA2^{het}$;$ApoE^{-/-}$ mice was visible after 8 and 12 weeks of HFD, indicating that not only does BRCA2 protect the vasculature against atherosclerosis, but also this role is conferred in relation to the number of functional *BRCA2* alleles in the endothelium. To our surprise, net weight, weight gain and lipid profiling displayed similar pattern among the genotypes post-HFD (Fig. 5E and Table 1) irrespective of the previously described association of *BRCA2* mutations and lipid dysregulation in humans.[61, 120] Based on a study by Miao et al., GG, GT and GT/GG genotypes of *BRCA2* single nucleotide polymorphism rs9534275 are associated with increased serum total cholesterol in humans.[59] However, in our previous studies, prevention[67] or exacerbation[65] of atherosclerosis did not significantly affect lipid and cholesterol levels in HFD-fed $ApoE^{-/-}$ mice. It is important to note that lipid and cholesterol dysregulation was reported in systemic heterozygous *BRCA2* mutant animal models,[59, 61, 120] whereas in the current study, *BRCA2* was knocked out only in endothelium. Therefore, it can be inferred that endothelial-specific BRCA2 haploinsufficiency does not result in lipid dysregulation.

Dissection and gross examination of the double knockout mice revealed remarkable splenomegaly and a conspicuous trend towards increased spleen weight was noticed in $BRCA2^{het}$;$ApoE^{-/-}$ and $BRCA2^{endo}$;$ApoE^{-/-}$ mice at 12 weeks post-HFD (Fig. 7B). HFD-induced splenomegaly was previously reported in $ApoE^{-/-}$ mice.[127] Molecular and histological analysis of the spleens confirmed HFD-induced splenic lymphoid hyperplasia. A relationship between splenomegaly and lipid metabolic disorders,[121] cardiac and endothelial dysfunction,[122, 123] systemic inflammation,[124] and atherosclerosis,[125, 126] is already established. BRCA2 is known to regulate HO-1 expression via binding to NRF2,[73, 74] and HO-1 is an important modulator of endothelial function.[128] Our data showed downregulation of HO-1 in the spleen of HFD-fed $BRCA2^{het}$;$ApoE^{-/-}$ and $BRCA2^{endo}$;$ApoE^{-/-}$ mice, which may be due to the inability of NRF2 to bind to BRCA2 that leads to inhibition of HO-1 transcription.

BRCA2 protein is one the largest eukaryotic proteins (384 kDa) with multiple domains such as BRC, DNA-binding domain (DBD), and nuclear localization sequence (NLS) which directly and indirectly interact with different protein partners throughout DSB repair.[129, 130] The BRCA2 DBD contains a helical domain (H), three oligonucleotide binding (OB) folds and a tower domain (T), which facilitate binding to both single-stranded and double-stranded DNA.[130, 131] Previous research on BRCA2 function was mainly focused on the mechanisms of DDR, however, additional investigation is warranted to further characterize different domains of BRCA2 and potential binding partners involved in pathways other than DDR, such as antioxidant pathway. Recently, it was proposed that BRCA2 regulates the NRF2 antioxidant pathway via PALB2,[73] as well as regulation of inflammation[39] and lipidmetabolism.[61, 120] In order to obtain a comprehensive insight regarding potential multiple roles of BRCA2 in the cardiovascular system, we performed RNA sequencing on the aortas of HFD-fed $BRCA2^{endo}$;$ApoE^{-/-}$ and $BRCA2^{WT}$;$ApoE^{-/-}$ mice and identified 530 DEGs associated with the atherosclerotic phenotype attributed to endothelial-specific loss of *BRCA2*. The most upregulated gene, *SYCP1*, is involved in formation of the synaptonemal complex by pairing homologous chromosomes during meiosis.[132] SYCP1 overexpression was reported in a variety of neoplasms including breast cancer.[133] Additionally, accelerated aging was reported in SYCP1-deficient mice[134] and in the women carrying germline *BRCA2* mutation in the mammary epithelia.[135] The most downregulated gene, *SNORD15B*, is a non-coding nucleolar RNA that modifies other small nuclear RNAs (snRNAs).[136] Here, for the first time we report a link between SNORD1B, SYCP1, and BRCA2 deficiency with atherosclerosis . IPA analysis identified Protein Folding and Cholesterol Biosynthetic Process as the most affected pathways. It is important to note that Protein Folding response is known to regulate apoptosis,[137] and its dysregulation can increase the susceptibility of the BRCA2-deficient endothelium to injury and an increased plaque. Analysis of upstream regulators highlighted the activation of key molecules associated with endothelial dysfunction. TGFBR2 activation is associated with endothelial to mesenchymal transition and development of atherosclerosis.[70, 138] Activation of SREBP2 was shown to increase oxidative stress in ECs[139] and promote atherosclerosis via activation of NLRP3 inflammasome.[140] Our results are in consistence with these findings suggesting that loss of *BRCA2* leads to disturbed molecular signalling that increases endothelial sensitivity under stressful environments such as hyperglycemia, genotoxic stress and hypercholesterolemia. The mouse genome dataset used for analysis of our RNA sequencing data provides comprehensive coverage of greater than 200k transcripts including both coding and non-coding transcripts such as long non-coding RNAs (lncRNAs), small RNAs, and those produced from pseudogenes. A major future challenge is to functionally annotate these genes and to ascertain their contribution, if any, to endothelial function and/or atherosclerosis.

Taken together, our results demonstrate an important role for endothelial BRCA2 shielding against plaque deposition and splenomegaly. Further mechanistic investigations are required to validate our findings from RNA sequencing analysis and broaden our understanding of BRCA2 function apart from its canonical contribution to DDR. Our RNA sequencing analyses warrants the necessity to explore beyond the role of BRCA2 in DDR and genome integrity maintenance in different CVDs. Additionally, our research creates a bridge between cancer and CVDs and provides strong evidence for correlation between BRCA2 and atherogenesis. This is a critical step towards verifying the existence of a BRCA2-atherosclerosis axis within human cardio-oncological landscape. Evaluation of *BRCA2* mutation status among breast cancer patients will identify those at increased risk of CVDs and may provide an invaluable tool for guiding precision treatments in these individuals.

**Author Contribution**

Conceptualization—KS; methodology—DCR, SN, BR, MP, SB, HCN, AS, BF, JM, RG, JF; formal analysis—DCR, KS, RG, ME.; data interpretation— DCR, KS, ME, RG, JF; writing — DCR, KS, SN, ME, JF. All authors have read and agreed to the published version of the manuscript.

**Sources of Funding**

Funding for this project was provided by the Project Grant (FRN # 153216), Canadian Institutes of Health Research, Canada to KS. KS is also the recipient of the 2018/19 National New Investigator Award- Salary Support from the Heart and Stroke Foundation of Canada, Canada.

## Tables

**Table 1.** Lipid profiling.

| Lipid parameters (mg/dL) | $ApoE^{-/-}$;$BRCA2^{WT}$ | $ApoE^{-/-}$;$BRCA2^{endo}$ | *p*-value |
|---|---|---|---|
| Total cholesterol | 1215.9±173.6 | 1371.4±116.5 | 0.18 |
| HDL cholesterol | 33.3±17.1 | 27.9±5.6 | 0.60 |
| Triglyceride | 199.9±65.3 | 193.9±18.9 | 0.87 |
| LDL cholesterol | 345.0±68.2 | 393.5±59.1 | 0.29 |
| Glucose | 261.2±25.0 | 251.6±43.1 | 0.64 |
| LDL calculated | 1133.0±150.7 | 1309.8±118.9 | 0.10 |

**Table 1.** Lipid parameters in male $ApoE^{-/-}$;$BRCA2^{WT}$ (n=6) and $ApoE^{-/-}$;$BRCA2^{endo}$ (n=4) mice after 12 weeks on a high-fat diet. Values are presented as mean ± SD. HDL: high-density lipoprotein; LDL: low-density lipoprotein.

**Table 2.** Top upregulated DEGs in the aorta of $BRCA2^{endo}$;$ApoE^{-/-}$ vs. $BRCA2^{WT}$;$ApoE^{-/-}$ mice.

| No. | Gene symbol | Fold-change | *p*-value |
|---|---|---|---|
| 1 | *Sycp1* | 415 | 3.92E-08 |
| 2 | *Gm45677* | 274 | 5.40E-06 |
| 3 | *Gm18562* | 102 | 7.08E-04 |
| 4 | *Gm23262* | 61.3 | 2.73E-03 |
| 5 | *Kap* | 58.4 | 3.18E-03 |
| 6 | *Gm42733* | 56.9 | 8.23E-04 |
| 7 | *Gm7645* | 52.4 | 1.70E-04 |
| 8 | *Iqschfp* | 50.3 | 6.55E-03 |
| 9 | *Gm18953* | 47.4 | 6.79E-05 |
| 10 | *Cbx3-ps7* | 36.1 | 1.32E-02 |
| 11 | *Aadat* | 32.5 | 4.85E-03 |
| 12 | *Mup15* | 30.5 | 2.26E-03 |
| 13 | *Gm6061* | 18.9 | 3.47E-02 |
| 14 | *Gm50367* | 17.9 | 5.16E-03 |
| 15 | *Gm29603* | 15.4 | 4.78E-02 |
| 16 | *Gm7328* | 15.1 | 4.01E-02 |
| 17 | *Ighv11-2* | 14.3 | 9.39E-05 |
| 18 | *Serpina4-ps1* | 13.0 | 4.95E-02 |
| 19 | *Slc34a1* | 11.0 | 2.00E-02 |
| 20 | *Mup14* | 10.7 | 1.34E-02 |

**Table 2.** Top upregulated differentially expressed genes (DEGs) in the aorta of $BRCA2^{endo}$;$ApoE^{-/-}$ vs. $BRCA2^{WT}$;$ApoE^{-/-}$ mice. *Aadat: aminoadipate aminotransferase; Cbx3-ps7: chromobox 3, pseudogene 7; Gm18562: ERO1-like beta pseudogene; Ighv11-2: immunoglobulin heavy variable V11-2; Iqschfp: Iqcj and Schip1 fusion protein; kap: kidney androgen regulated protein; Mup14: major urinary protein 14; Mup15: major urinary protein 15; Serpina4-ps1: serine (or cysteine) peptidase inhibitor, clade A, member 4, pseudogene 1; Slc34A1: solute carrier family 34 member*

*1; SYCP1: synaptonemal complex protein 1*. The remaining gene symbols starting with '*Gm*' refer to predicted genes.

**Table 3.** Top downregulated DEGs in the aorta of BRCA2$^{endo}$;ApoE$^{-/-}$ vs. BRCA2$^{WT}$;ApoE$^{-/-}$ mice.

| No. | Gene symbol | Fold-change | *p*-value |
|---|---|---|---|
| 1 | *Snord15b* | -193 | 8.95E-10 |
| 2 | *Gm49739* | -7.79 | 3.66E-03 |
| 3 | *Eif1ad2* | -39.2 | 2.69E-03 |
| 4 | *Zxda* | -34.1 | 4.42E-02 |
| 5 | *Gm6741* | -30.4 | 4.70E-02 |
| 6 | *H3f3c* | -30.0 | 2.42E-03 |
| 7 | *Gm29542* | -27.3 | 3.27E-02 |
| 8 | *Septin14* | -22.2 | 1.02E-03 |
| 9 | *Krtap3-2* | -20.3 | 4.18E-02 |
| 10 | *Atp5o* | -15.4 | 1.90E-03 |
| 11 | *Gm19951* | -13.9 | 5.24E-03 |
| 12 | *Gp1bb* | -13.7 | 1.33E-03 |
| 13 | *Pagr1b* | -12.3 | 2.28E-03 |
| 14 | *Gm16477* | -11.4 | 2.96E-02 |
| 15 | *Gm15387* | -11.1 | 1.62E-02 |
| 16 | *H1f10* | -10.7 | 5.48E-03 |
| 17 | *Gm35082* | -10.0 | 2.74E-02 |
| 18 | *Trim34b* | -9.23 | 2.77E-05 |
| 19 | *Htr2c* | -7.90 | 4.77E-02 |
| 20 | *Gm10447* | -7.46 | 1.07E-02 |

**Table 3.** Top down-regulated differentially expressed genes (DEGs) in the aorta of BRCA2$^{endo}$;ApoE$^{-/-}$ vs. BRCA2$^{WT}$;ApoE$^{-/-}$ mice. *Snord15b: small Nucleolar RNA, C/D Box 15B; Atp5o: ATP synthase, H+ transporting, mitochondrial F1 complex, O subunit; Eif1ad2: eukaryotic translation initiation factor 1A domain containing 2; Pagr1b: PAXIP1 associated glutamate rich protein 1B; Gm1538: high mobility group protein 1 pseudogene; Gm16477: ribosomal protein L7a pseudogene; Gp1bb: glycoprotein Ib platelet subunit beta; H1f10: H1.10 linker histone; H3f3c: H3 histone, family 3C; Htr2c: 5-hydroxytryptamine receptor 2C; Krtap3-2: keratin*

*associated protein 3-2; Trim34b: tripartite motif-containing 34B; Zxda: zinc finger X-linked duplicated A*. The remaining gene symbols starting with '*Gm*' refer to predicted genes.

## Figure Legends

**Figure 1. Baseline BRCA2 expression in different organs of mice and EC-specific Cre-mediated deletion of BRCA2. (A)** RT-qPCR was performed on RNA extracted from ovary, testis, liver, spleen, left and right ventricle, atria and aorta of WT mice. Bars represent the expression of BRCA2 transcript in different organs. BRCA2 expression in ovary was considered as 100%. All organs except ovary were isolated from 8-10 weeks old male mice. n=3-4/group. **(B)** Breeding strategy used for the generation of $BRCA2^{endo}$, $BRCA2^{het}$ and WT (Cre-only; control) mice using the Cre-loxP technology. Briefly, mice homozygous for the exon 11 floxed *BRCA2* allele ($BRCA2^{fl/fl}$) were crossed with hemizygous mice expressing Cre recombinase under the control of the endothelial cell-specific VE-cadherin promoter (VE-cadherin-$Cre^{tg/-}$). Mice carrying VE-cadherin-$Cre^{tg/-}$;$BRCA2^{fl/wt}$ combination were denoted as EC-specific BRCA2 heterozygous ($BRCA2^{het}$) and mice carrying VE-cadherin-$Cre^{tg/-}$;$BRCA2^{fl/fl}$ combination were denoted as EC-specific BRCA2 homozygous ($BRCA2^{endo}$) knockdown and mice carrying VE-cadherin-$Cre^{tg/-}$;$BRCA2^{wt/wt}$ combination were denoted as Cre-only WT ($BRCA2^{WT}$). **(C)** Schematic representation of Cre-mediated deletion of exon 11 of *BRCA2* gene. Arrows denote the binding sites of primers for genotypic identification. **(D)** Representative PCR products of genomic DNA isolated from the ear notches of mice using primer pair specific to the 5' loxP site. PCR products of 376bp and 529bp for primers 12/13 and 14/15, respectively, confirms the presence of the loxP site. Presence of VE-cadherin-controlled Cre is identified by the amplification of a 100 bp sequence with primer pair oIMR1084/1085 and second primer pair oIMR7338/7339amplifies a 324 bp sequence as a positive control.

**Figure 2. Successful generation of EC-specific BRCA2 knockdown Mice. (A)** PCR amplification of BRCA2 exon 11 in lung RNA extract demonstrating the genotype-dependent

reduction of exon 11 in BRCA2$^{endo}$ and BRCA2$^{het}$ mice relative to their BRCA2$^{WT}$ counterparts. **(B)** Immunoblotting on proteins extracted from lung tissue lysate with β-tubulin as a loading control. **(C)** Histological images of BRCA2 antibody-incubated DAB-stained aorta and lung of BRCA2$^{WT}$, BRCA2$^{het}$, and BRCA2$^{endo}$ mice. **(D)** Agarose gel electrophoresis image exhibiting the truncated BRCA2 (905 bp) PCR product produced by BRCA2 exon 11 deletion only present in the aortic ECs of BRCA2$^{endo}$ mice. **(E)** Immunoblotting for BRCA2 with β-tubulin on total protein extracted from aortic ECs of BRCA2$^{WT}$ and BRCA2$^{endo}$ mice. All experiments were done using 8-10 weeks old male mice. n=3/group.

**Figure 3. Similar weight, DNA damage and repair, and apoptosis at baseline in BRCA2$^{endo}$ mice.** **(A)** Mendelian ratio and **(B)** mice weight was obtained at 8-10 weeks of age (n=20-30/group). Next, aorta, lung, and liver were dissected from perfused BRCA2$^{endo}$, BRCA2$^{het}$, and BRCA2$^{WT}$ littermates. Tissues were cut at 5 microns and sections were stained with antibodies for **(C)** γH2AX . Nuclei are stained blue by hematoxylin and antibody signal is visible as brown precipitate formed by horseradish peroxidase (HRP)-catalyzed Diaminobenzidine (DAB). **(D)** Proteins were extracted from lung of BRCA2$^{WT}$ and BRCA2$^{endo}$ mice and immunoblot for γH2AX, p53 and GAPDH (loading control) was performed. **(E, F)** Representative terminal deoxynucleotidyl transferase dUTP Nick-End Labeling (TUNEL), DAPI counterstained nuclei, and colorized merged images of the aorta and lung showed no apoptosis in the endothelium of BRCA2$^{WT}$, BRCA2$^{het}$, or BRCA2$^{endo}$ mice. All experiments were done using 8-10 weeks old male mice. n=3/group.

**Figure 4. Similar cardiac function at baseline in BRCA2$^{WT}$, BRCA2$^{het}$ BRCA2$^{endo}$ mice.** **(A)** B-mode echocardiogram delineating the cardiac region displaying left ventricle end-diastolic diameter (LVEDD) and left ventricle end-systolic diameter (LVESD). **(B, C)** Similar ejection

fraction and fractional shortening were observed for BRCA2$^{WT}$, BRCA2$^{het}$ BRCA2$^{endo}$ mice at baseline. Statistical analysis was performed by one-way ANOVA with Tukey's post-hoc. Data are presented as mean ± SD. n=6/group.

**Figure 5. Generation of EC-specific BRCA2 knockout mice on ApoE$^{-/-}$ background.** **(A)** The BRCA2$^{endo}$ (VE-cadherin-Cre$^{tg/-}$;BRCA2$^{fl/fl}$:ApoE$^{wt/wt}$) mice were crossed with ApoE$^{-/-}$ mice to generate VE-cadherin-Cre$^{tg/-}$;BRCA2$^{fl/wt}$:ApoE$^{wt/-}$ mice. Later, VE-cadherin-Cre$^{tg/-}$;BRCA2$^{fl/wt}$:ApoE$^{wt/-}$ mice were crossed with BRCA2$^{fl/wt}$:ApoE$^{wt/-}$ to generate VE-cadherin-Cre$^{tg/-}$;BRCA2$^{fl/fl}$:ApoE$^{-/-}$ or BRCA2$^{endo}$:ApoE$^{-/-}$ mice. **(B)** Representative agarose gel electrophoresis image of genotype variants of ApoE mutation expression amplified by PCR. **(C)** Mice were born in expected Mendelian Ratio. Chi-Squared analysis demonstrated no statistical deviation from expected ratio (male n=50, female n=21, total n=71). **(D)** Mice displayed similar weight at baseline [10 weeks age, males, VE-cadherin-Cre$^{tg/-}$;BRCA2$^{wt/wt}$:ApoE$^{-/-}$ (n=4), BRCA2$^{fl/fl}$:ApoE$^{-/-}$ (n=40), VE-cadherin-Cre$^{tg/-}$;BRCA2$^{fl/wt}$:ApoE$^{-/-}$ (n=19), and VE-cadherin-Cre$^{tg/-}$;BRCA2$^{fl/fl}$:ApoE$^{-/-}$ (n=29)]. **(E)** Mice displayed similar weight following 4, 6, 8, 10 and 12 weeks of HFD. n=10-12/group.

**Figure 6. EC-specific loss of BRCA2 exacerbates atherosclerotic plaque burden.** **(A)** Representative en face imaged aortas of male mice fed HFD for 8 weeks demonstrating increased atherosclerotic plaque burden in the aortic arch of heterozygous and homozygous EC-specific BRCA2 knockout mice. Aortic plaque quantification in the aorta of 8 **(B)**, 12 **(C)** and 16 **(E)** weeks post-HFD (males, n=5-7/group). **(D)** Representative Oil Red-O-stained whole aortas of 12-week HFD-fed male mice. **(F)** Aortic roots were collected from BRCA2$^{WT}$;ApoE$^{-/-}$ and BRCA2$^{endo}$;ApoE$^{-/-}$ mice. Aortic roots were sectioned and stained for F4/80 and then quantified (males, n=4-6/group). Data are analyzed by one-way ANOVA with Tukey's post hoc (B, C, E) or student's T-test (G)

and are presented as mean ± SD. *, **, *** represents $p < 0.05$, 0.01 or 0.001 vs. BRCA2$^{WT}$;ApoE$^{-/-}$.

**Figure 7. EC-specific loss of BRCA2 causes HFD-induced splenomegaly in atherosclerotic mice. (A, B)** Spleen weight (SW)/Body weight (BW) were measured in BRCA2$^{WT}$;ApoE$^{-/-}$, BRCA2$^{het}$;ApoE$^{-/-}$ and BRCA2$^{endo}$;ApoE$^{-/-}$ mice 8 and 12-weeks post-HFD (males, n=4-9/group). **(C)** Spleens were collected from BRCA2$^{WT}$;ApoE$^{-/-}$, BRCA2$^{het}$;ApoE$^{-/-}$ and BRCA2$^{endo}$;ApoE$^{-/-}$ mice 12-weeks post-HFD, then sectioned and H&E staining was performed. **(D)** Total RNAs were collected from the spleens of 12-weeks HFD-fed BRCA2$^{WT}$;ApoE$^{-/-}$, BRCA2$^{het}$;ApoE$^{-/-}$ and BRCA2$^{endo}$;ApoE$^{-/-}$ mice, and then qPCR for HO-1 was performed (males, n=4-6/group). Data are analyzed by one-way ANOVA with Tukey's post hoc and are presented as mean ± SD. *, **represents $p < 0.05$ and 0.01 vs. BRCA2$^{WT}$;ApoE$^{-/-}$.

**Figure 8. RNA sequencing analysis on aortic RNAs obtained from BRCA2$^{WT}$;ApoE$^{-/-}$ and BRCA2$^{endo}$;ApoE$^{-/-}$ mice 12-weeks post-HFD. (A)** The range of quantitative change of up- and downregulated differentially expressed genes (DEGs) are shown in the Volcano Plot. Volcano plot showing log2 fold-change in gene expression in BRCA2$^{endo}$ compared to BRCA2$^{WT}$. Each dot represents a single gene. Red dots indicate DEGs ($p < 0.05$, 1.5 < fold-change > -1.5) **(B)** DE genes were annotated in the Ingenuity Pathway Analysis (IPA) database, which identified enrichment of multiple canonical pathways represented by DE genes. **(C)** IPA was used to predict the activation state of molecules. **(D)** We further categorized the 530 DE genes based on gene ontology (GO) by using the Database for Annotation, Visualization and Integrated Discovery (DAVID) into molecular function, biological process and cellular components. (n=4/group).

FIGURE 1

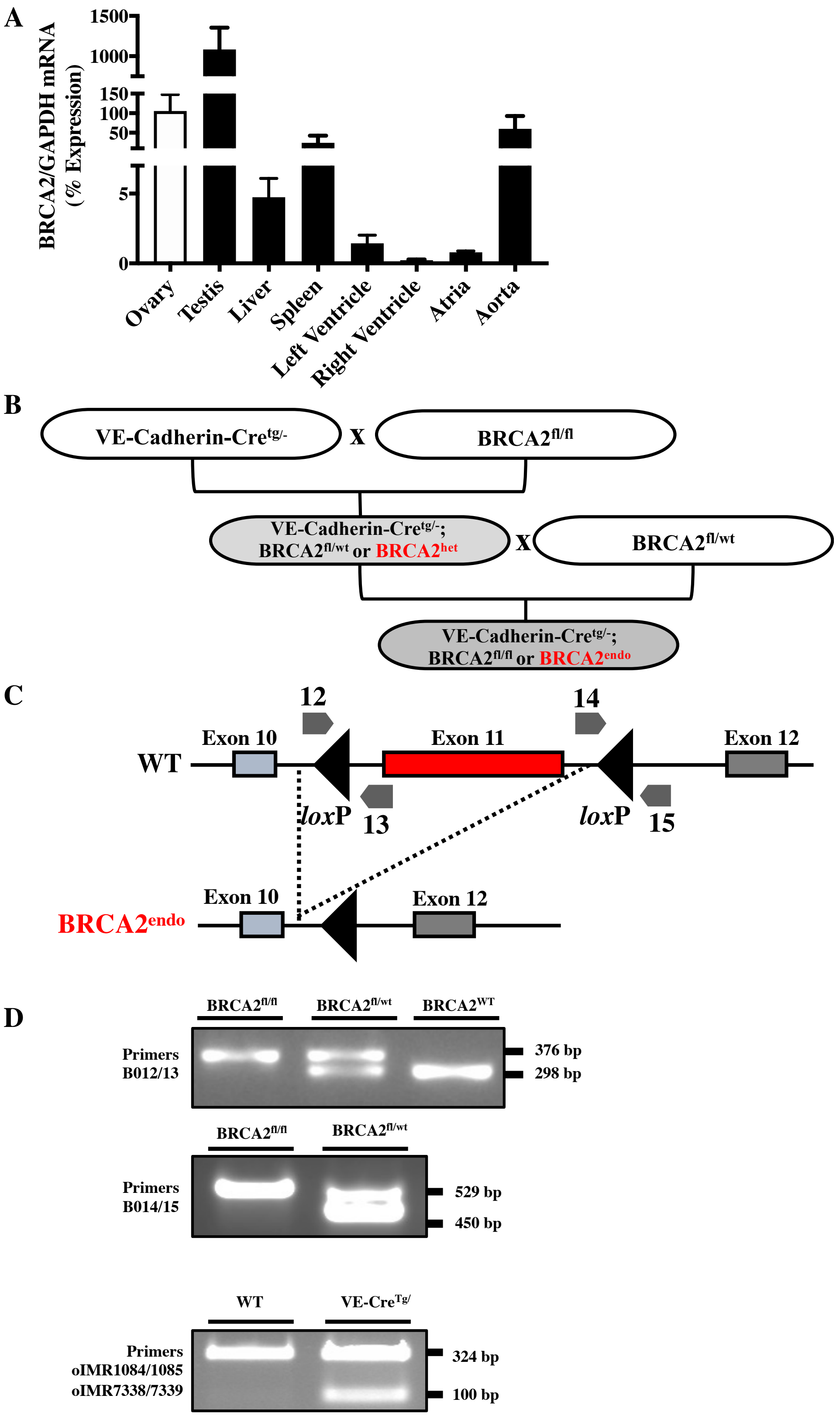
A
BRCA2/GAPDH mRNA
(% Expression)
1500
1000
150
100
50
5
0
Ovary
Testis
Liver
Spleen
Left Ventricle
Right Ventricle
Atria
Aorta
B
VE-Cadherin-Cre$^{tg/-}$
X
BRCA2$^{fl/fl}$
VE-Cadherin-Cre$^{tg/-}$;
BRCA2$^{fl/wt}$ or BRCA2$^{het}$
X
BRCA2$^{fl/wt}$
VE-Cadherin-Cre$^{tg/-}$;
BRCA2$^{fl/fl}$ or BRCA2$^{endo}$
C
12
14
Exon 10
Exon 11
Exon 12
WT
loxP
13
loxP
15
BRCA2$^{endo}$
Exon 10
Exon 12
D
BRCA2$^{fl/fl}$
BRCA2$^{fl/wt}$
BRCA2$^{WT}$
Primers
B012/13
376 bp
298 bp
BRCA2$^{fl/fl}$
BRCA2$^{fl/wt}$
Primers
B014/15
529 bp
450 bp
WT
VE-Cre$^{Tg/}$
Primers
oIMR1084/1085
oIMR7338/7339
324 bp
100 bp

FIGURE 2

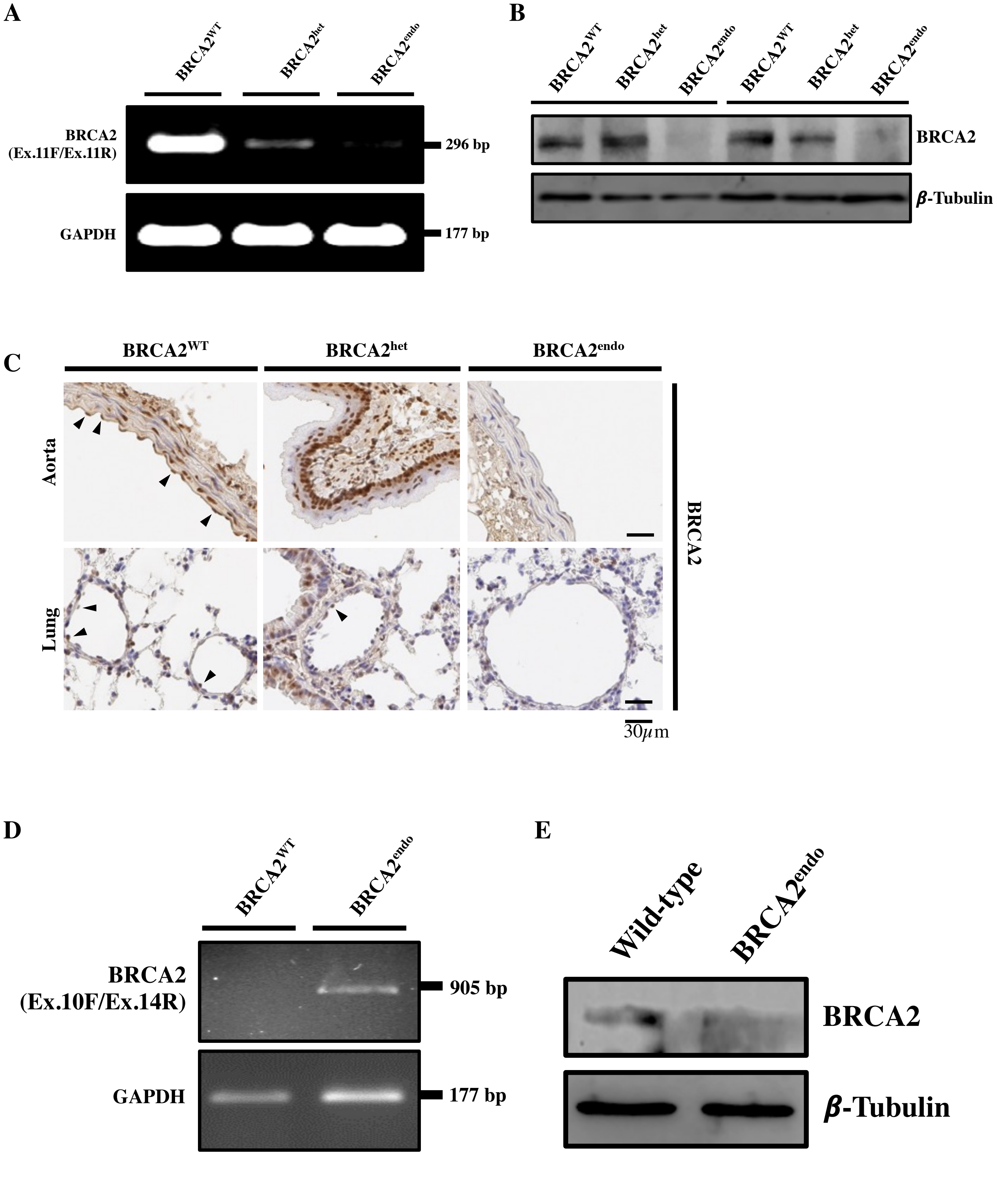

A

**FIGURE 3**

| VE-Cadherin-Cre$^{tg/}$;BRCA2$^{fl/wt}$ X BRCA2$^{fl/wt}$ | | | | | | |
|---|---|---|---|---|---|---|
| **Genotypes** | **BRCA2$^{wt/wtl}$** | **BRCA2$^{fl/wt}$** | **BRCA2$^{fl/fl}$** | **VE-Cadherin-Cre$^{tg/}$; BRCA2$^{wt/wt}$** | **VE-Cadherin-Cre$^{tg/}$; BRCA2$^{fl/wt}$** | **VE-Cadherin-Cre$^{tg/}$; BRCA2$^{fl/fl}$** |
| **Expected (%)** | 8.75 (12.5%) | 17.5 (25%) | 8.75 (12.5%) | 8.75 (12.5%) | 17.5 (25%) | 8.75 (12.5%) |
| **Observed (%)** | 5 (7.1%) | 21 (30%) | 5 (7.1%) | 11 (15.7%) | 20 (28.6%) | 8 (11.4%) |

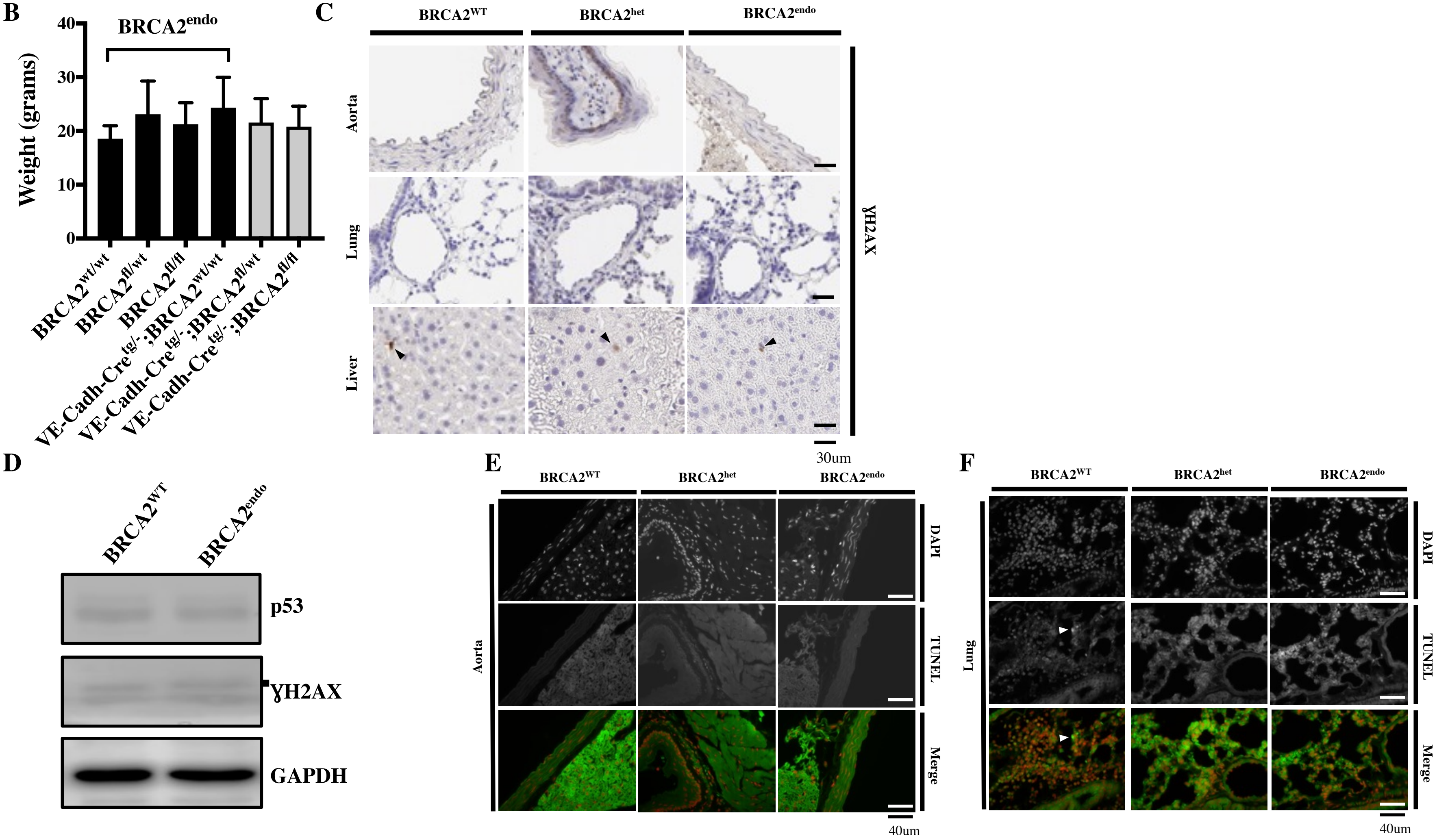

**FIGURE 4**

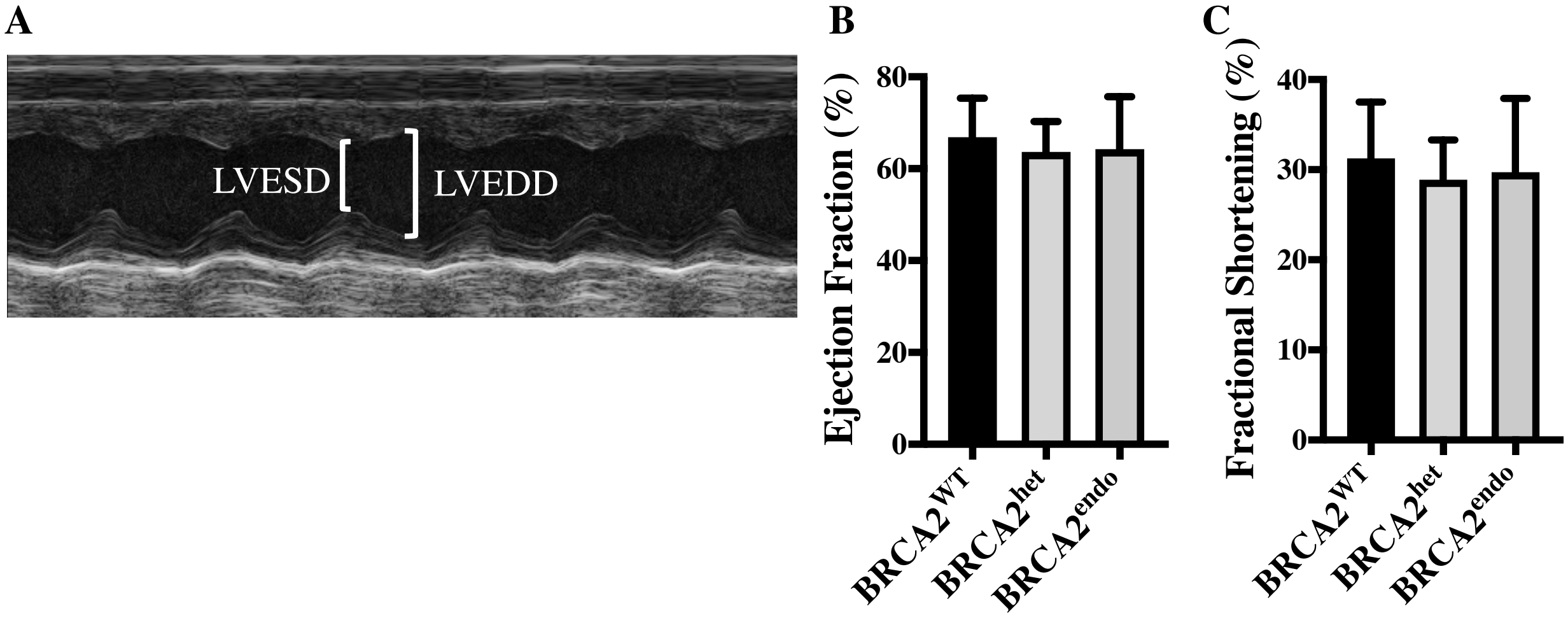

FIGURE 5

A

B

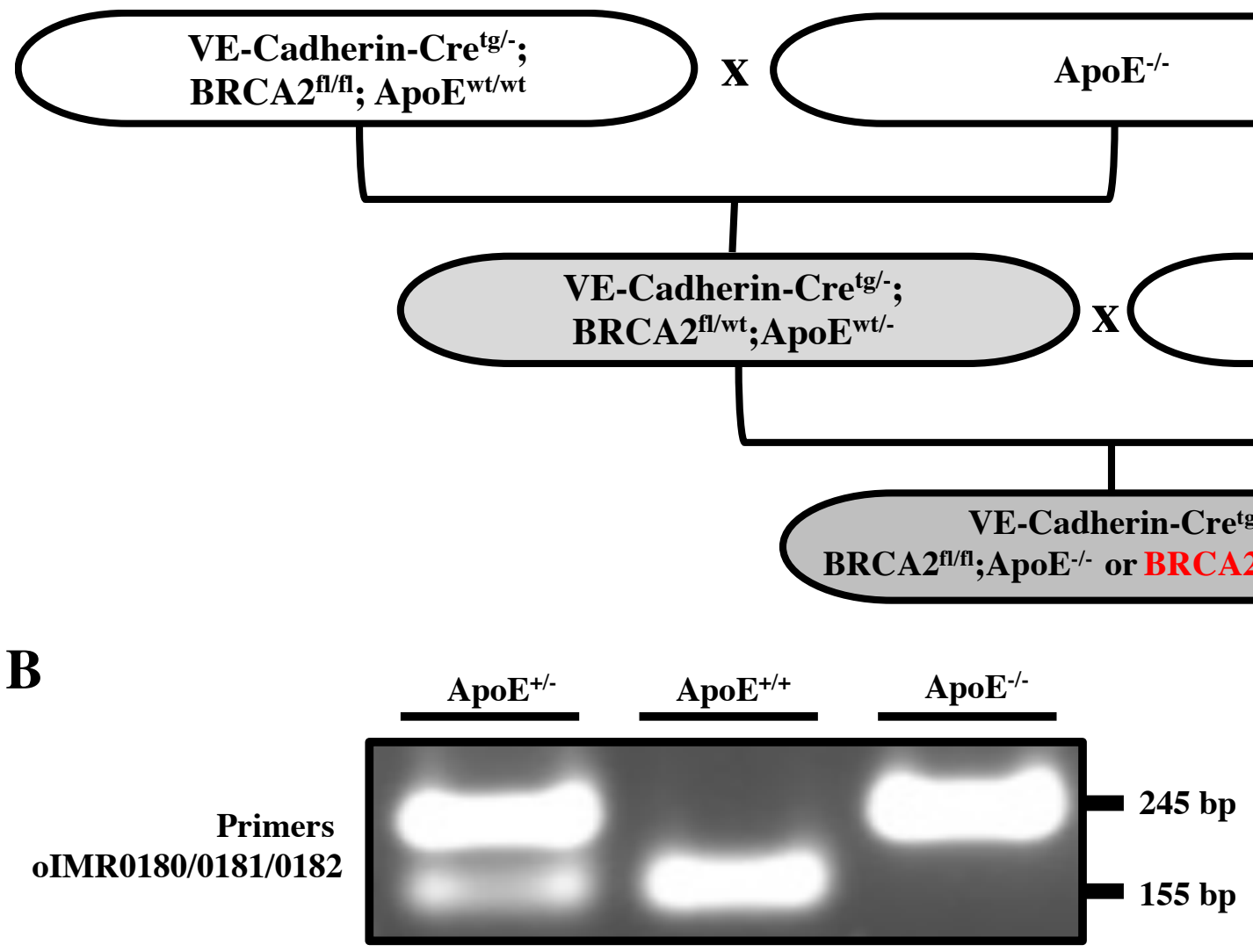

C

| VE-Cadherin-Cre$^{tg/-}$;BRCA2$^{fl/wt}$ X ApoE$^{-/-}$;BRCA2$^{fl/wt}$ | | | | |
|---|---|---|---|---|
| **Genotypes** | **VE-Cadherin-Cre$^{tg/-}$;BRCA2$^{fl/wt}$;ApoE$^{-/-}$** | **BRCA2$^{fl/wt}$;ApoE$^{-/-}$** | **VE-Cadherin-Cre$^{tg/-}$;BRCA2$^{fl/fl}$;ApoE$^{-/-}$** | **BRCA2$^{fl/fl}$;ApoE$^{-/-}$** |
| **Expected (%)** | 17.75 (25%) | 17.75 (25%) | 17.75 (25%) | 17.75 (25%) |
| **Observed (%)** | 19 (26.8%) | 19 (26.8%) | 12 (16.9%) | 21 (29.6%) |

D

E

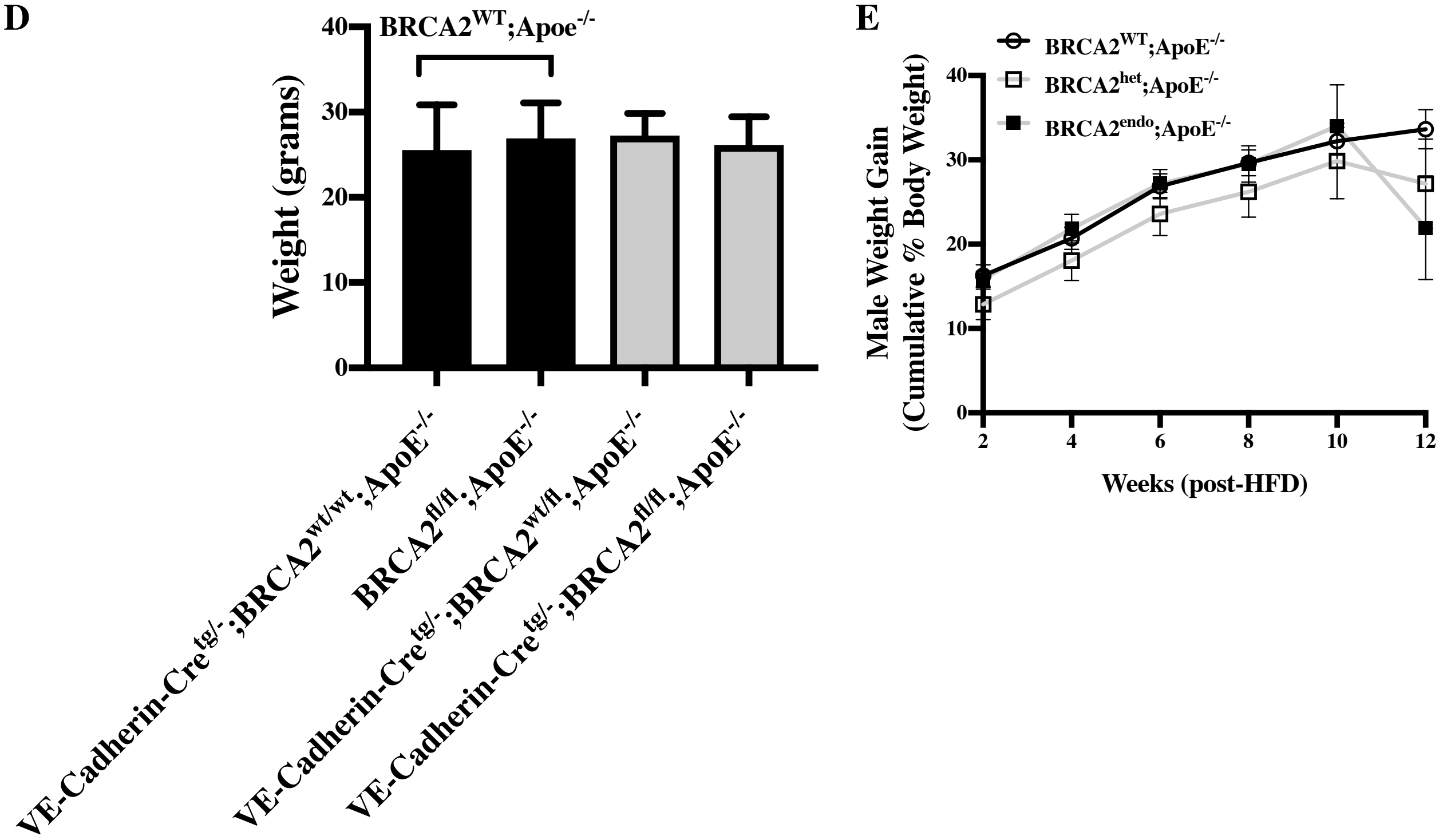

**FIGURE 6**

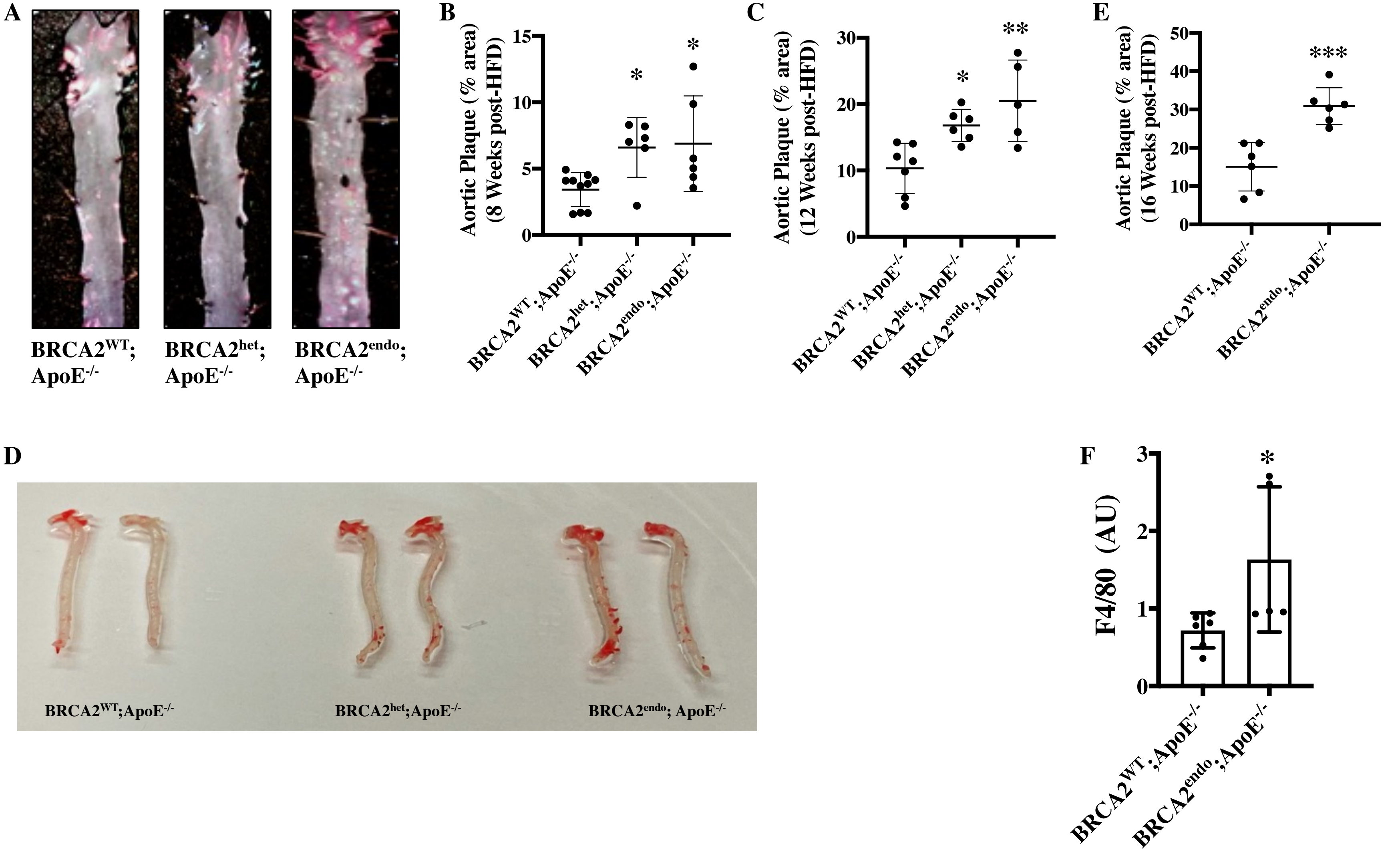

**FIGURE 7**

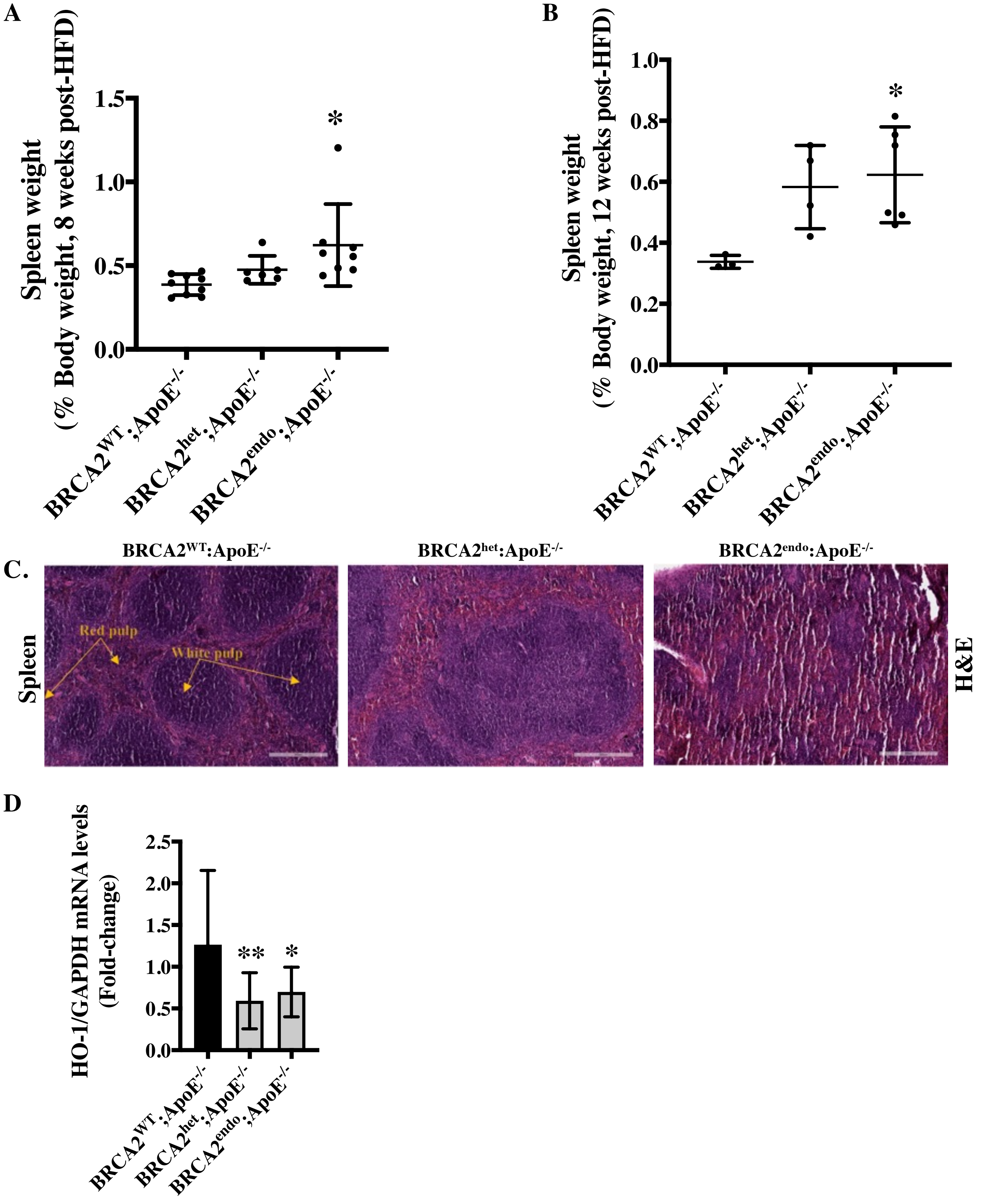

**FIGURE 8**

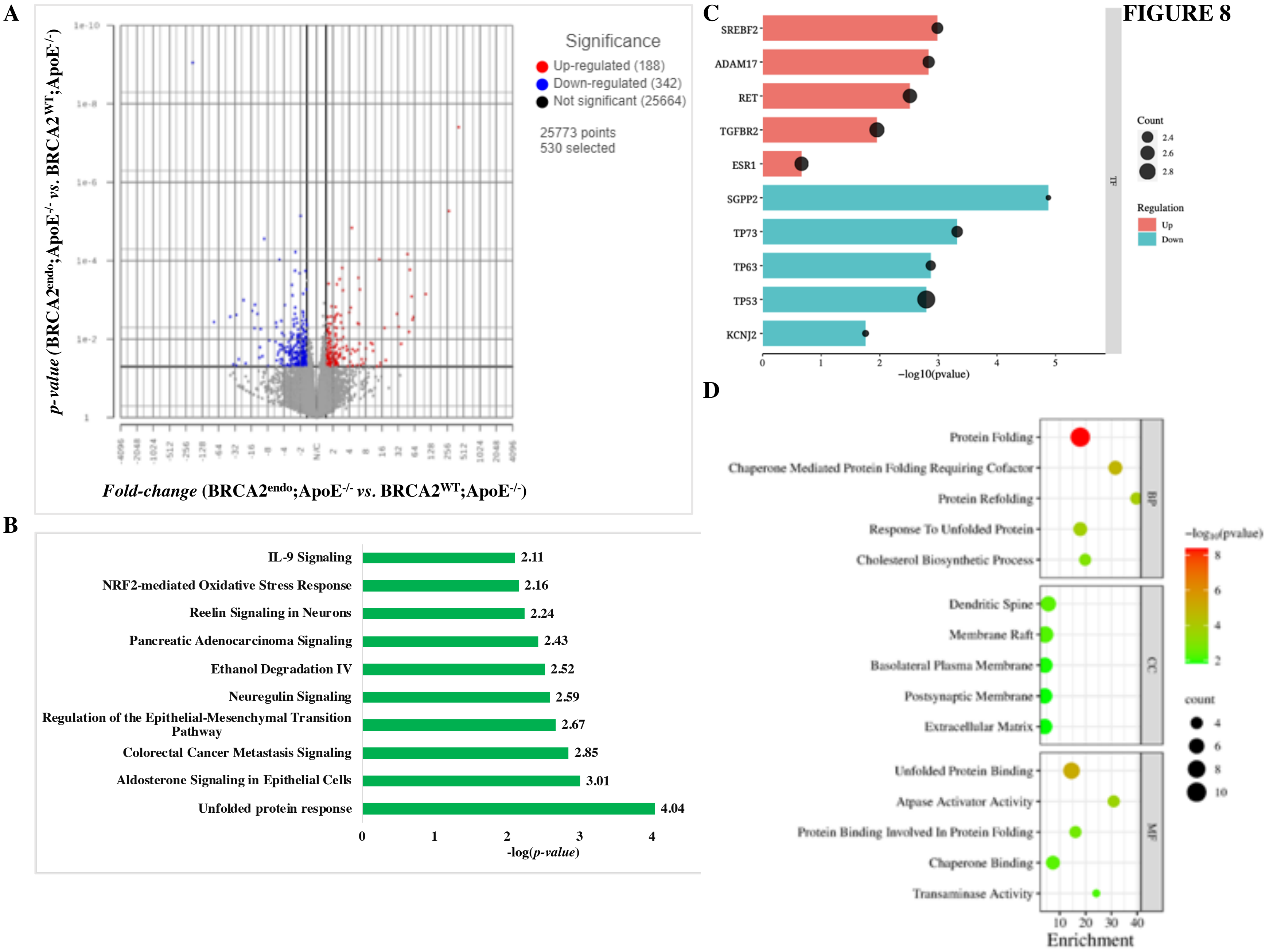
A
Significance
Up-regulated (188)
Down-regulated (342)
Not significant (25664)
25773 points
530 selected
p-value (BRCA2endo;ApoE-/- vs. BRCA2WT;ApoE-/-)
Fold-change (BRCA2endo;ApoE-/- vs. BRCA2WT;ApoE-/-)
B
IL-9 Signaling 2.11
NRF2-mediated Oxidative Stress Response 2.16
Reelin Signaling in Neurons 2.24
Pancreatic Adenocarcinoma Signaling 2.43
Ethanol Degradation IV 2.52
Neuregulin Signaling 2.59
Regulation of the Epithelial-Mesenchymal Transition Pathway 2.67
Colorectal Cancer Metastasis Signaling 2.85
Aldosterone Signaling in Epithelial Cells 3.01
Unfolded protein response 4.04
0 1 2 3 4
-log(p-value)
C
SREBF2
ADAM17
RET
TGFBR2
ESR1
SGPP2
TP73
TP63
TP53
KCNJ2
−log10(pvalue)
Count
2.4
2.6
2.8
Regulation
Up
Down
D
Protein Folding
Chaperone Mediated Protein Folding Requiring Cofactor
Protein Refolding
Response To Unfolded Protein
Cholesterol Biosynthetic Process
Dendritic Spine
Membrane Raft
Basolateral Plasma Membrane
Postsynaptic Membrane
Extracellular Matrix
Unfolded Protein Binding
Atpase Activator Activity
Protein Binding Involved In Protein Folding
Chaperone Binding
Transaminase Activity
BP
CC
MF
count
Enrichment

**Supplementary Figure 1**

**A**

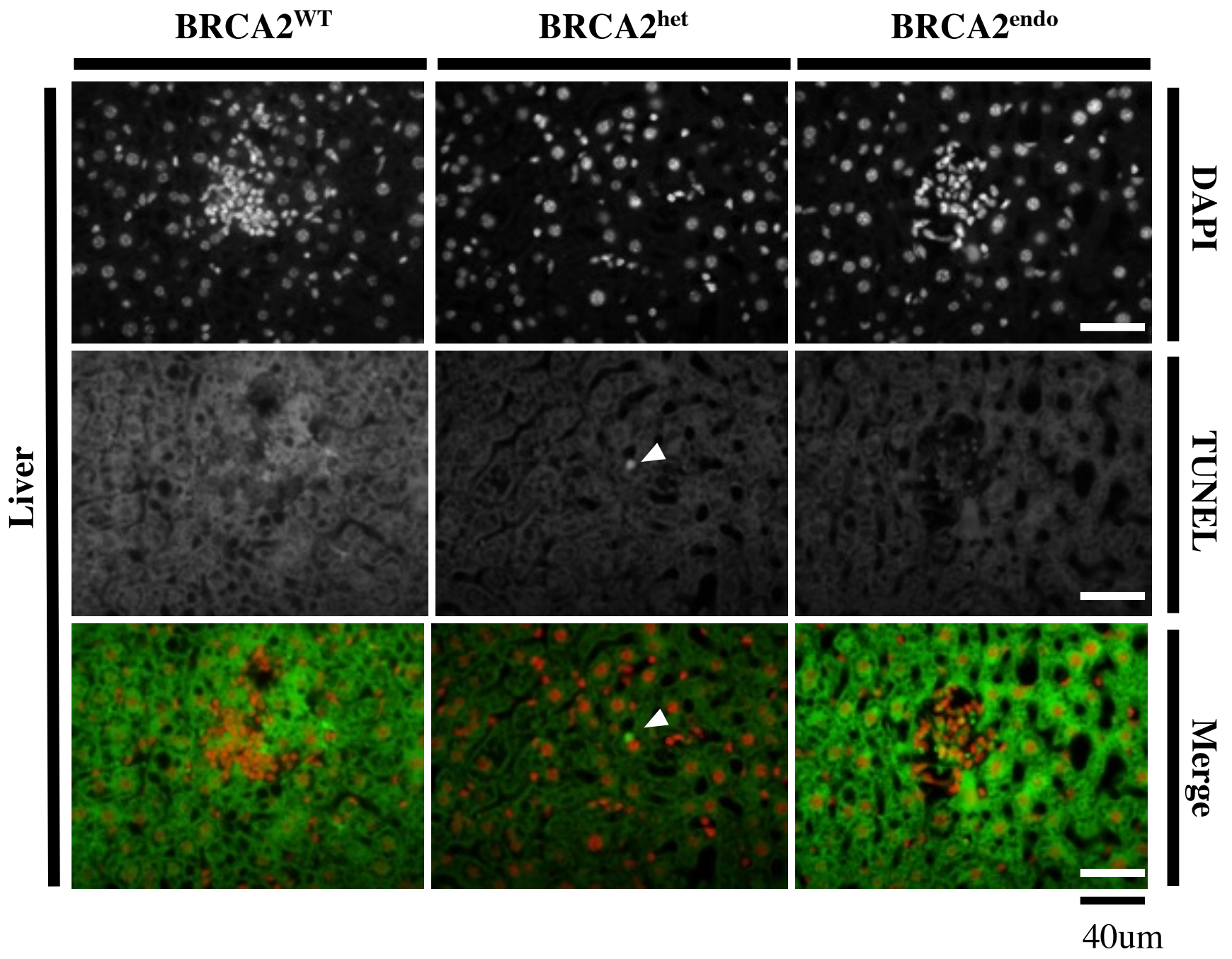

# SUPPLEMENTARY FIGURE 2

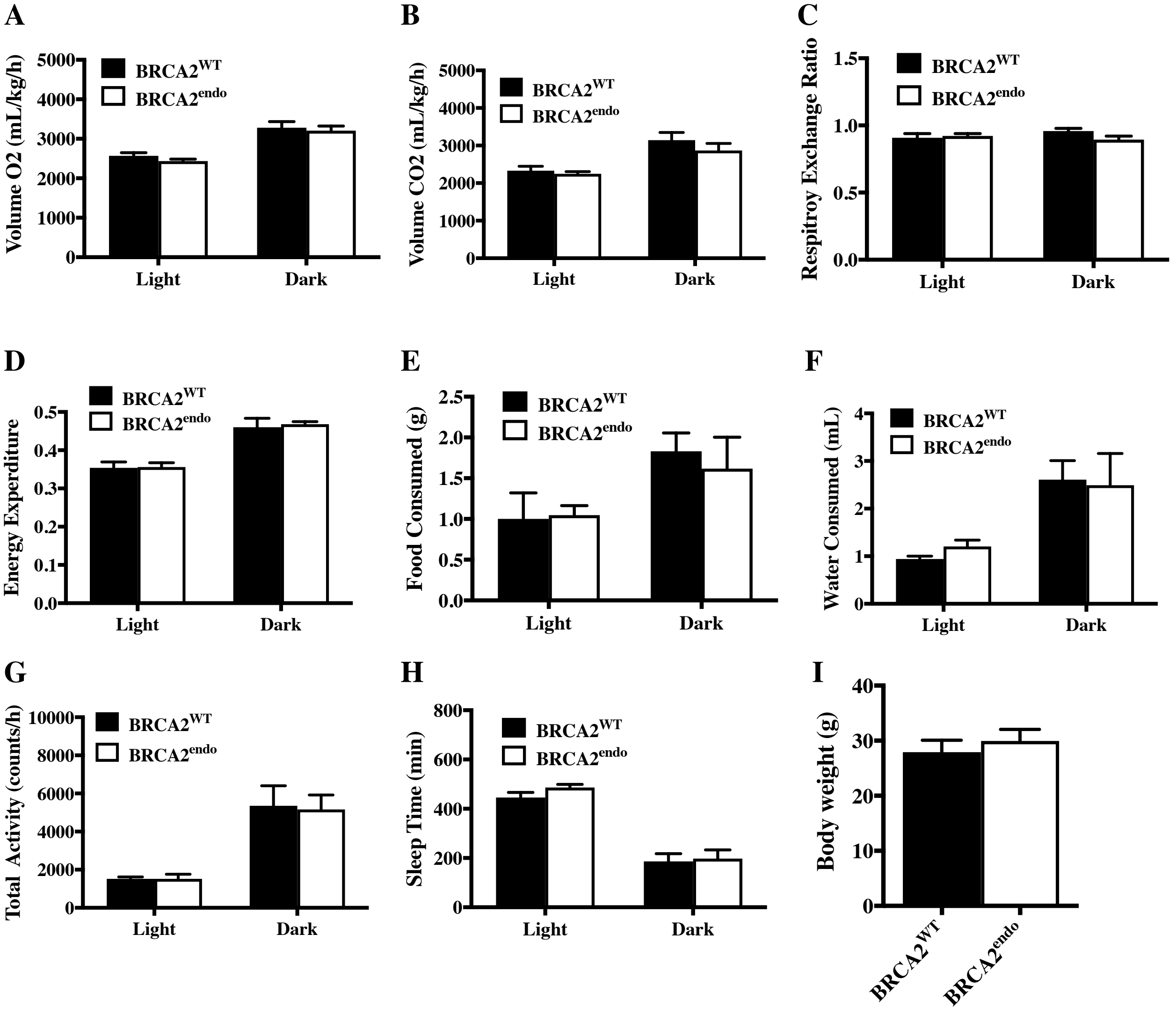

SUPPLEMENTARY FIGURE 3

A

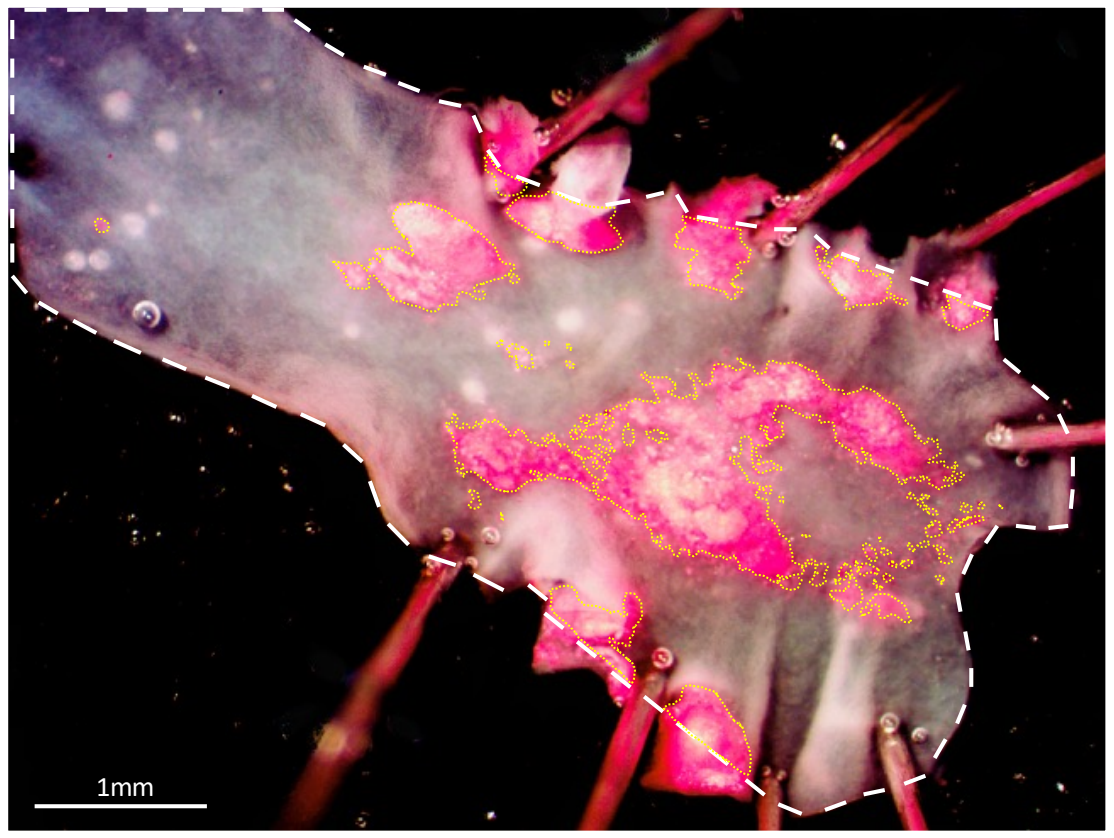

B

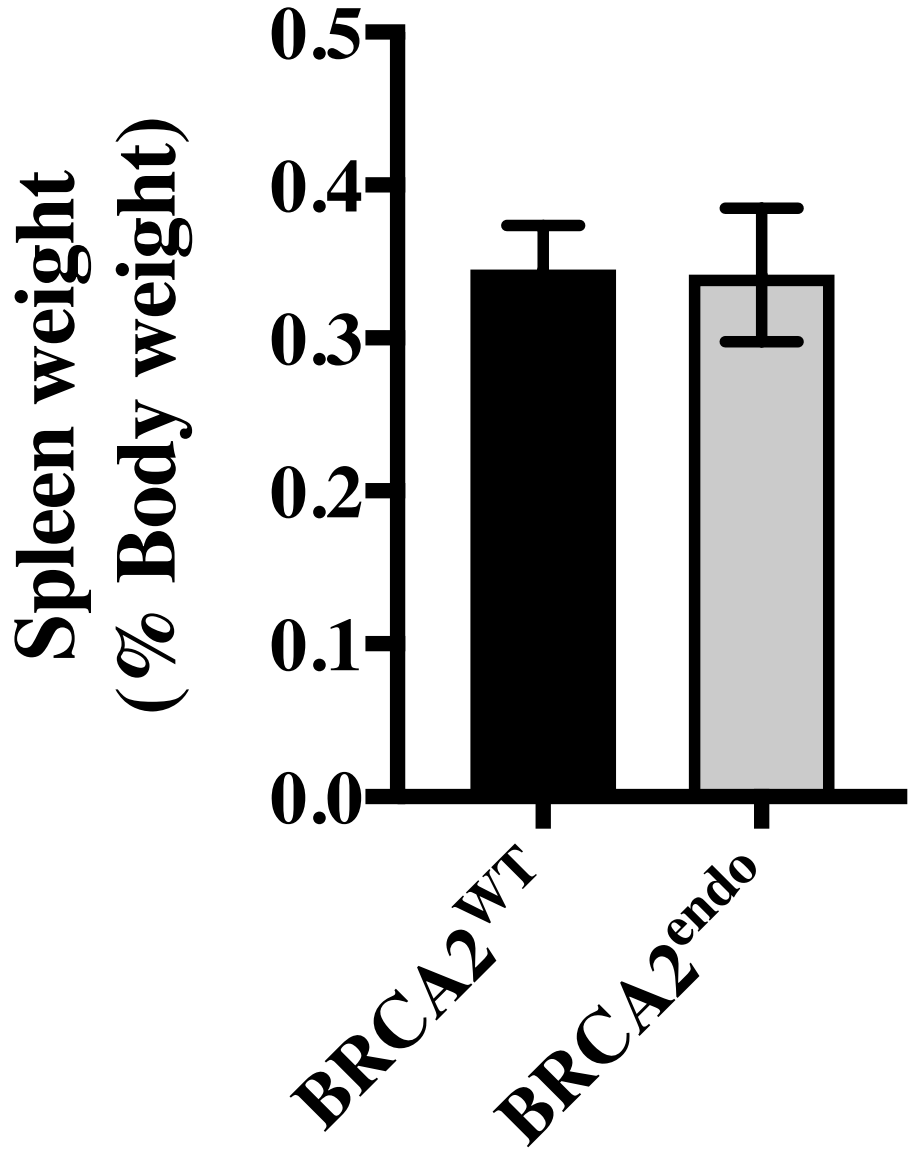

C

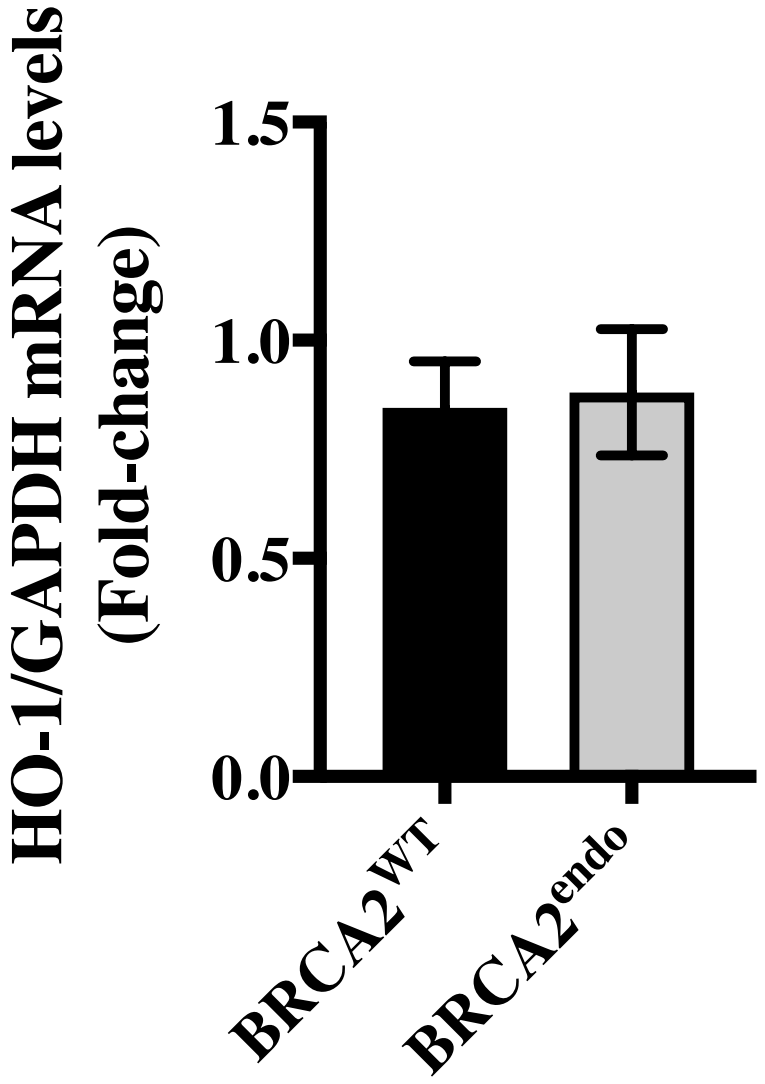